\documentclass[aps,prb,twocolumn,showpacs,superscriptaddress,groupedaddress,floatfix]{revtex4}

\usepackage{amsmath}
\usepackage{graphicx}

\newcommand{\mub}{$\mu_{\rm B}$}
\newcommand{\TN}{T$_{\rm N}$}

\newcommand{\Moss}{M\"ossbauer}
\newcommand{\xmu}{$x_\mu$}
\newcommand{\xmub}{$\bar{x}_\mu$}

\begin{document}
\author{N. Martin}
\affiliation{Laboratoire L\'eon Brillouin, CEA, CNRS, Universit\'e Paris-Saclay, CEA Saclay 91191 Gif-sur-Yvette, France}
\author{M. Deutsch}
\affiliation{Universit\'e de Lorraine, Laboratoire CRM2,UMR UL-CNRS 7036,
54506 Vandoeuvre-les-Nancy,France}
\author{F. Bert}
\affiliation{Laboratoire de Physique du Solide, UMR CNRS 8502, Universit\'e Paris Sud, FR-91140 Orsay  France}
\author{D. Andreica}
\affiliation{Faculty of Physics, Babes-Bolyai University, 400084 Cluj-Napoca, Romania}
\author{A. Amato}
\affiliation{Laboratory for Muon Spin Spectroscopy,  PSI, CH-5232 Villigen PSI, Switzerland}
\author{P. Bonf\`a}
\affiliation{Dipartimento di Fisica e Scienze della Terra and Unit\`a CNISM di Parma, Universit\`a di Parma, 43124 Parma, Italy}
\author{R. De Renzi}
\affiliation{Dipartimento di Fisica e Scienze della Terra and Unit\`a CNISM di Parma, Universit\`a di Parma, 43124 Parma, Italy}
\author{U.K. R\"{o}{\ss}ler}
\affiliation{Leibnitz Institute for Solid State and Material Research IFW Dresden, D-01069, Germany}
\author{P. Bonville}
\affiliation{SPEC, CEA, CNRS, Universit\'e Paris-Saclay, CEA-Saclay, 91191 Gif-sur-Yvette, France}
\author{L. N. Fomicheva}
\affiliation{Vereshchagin  Institute for  High Pressure Physics, Russian Academy of Sciences, 142190, Troitsk, Moscow, Russia}
\author{A. V. Tsvyashchenko}
\affiliation{Vereshchagin  Institute for  High Pressure Physics, Russian Academy of Sciences, 142190, Troitsk, Moscow, Russia}
\affiliation{Skobeltsyn Institute of Nuclear Physics, MSU, Vorob'evy Gory 1/2, 119991 Moscow, Russia}
\author{I. Mirebeau}
\email[E-mail: ]{isabelle.mirebeau@cea.fr}
\affiliation{Laboratoire L\'eon Brillouin, CEA, CNRS, Universit\'e Paris-Saclay, CEA Saclay 91191 Gif-sur-Yvette, France}

\title{Magnetic ground state and spin fluctuations in MnGe chiral magnet as studied by Muon Spin Rotation}

\date{\today}

\begin{abstract}
We have studied by  muon spin resonance ($\mu$SR) the helical ground state and fluctuating chiral phase recently observed in the MnGe chiral magnet. At low temperature, the muon polarization shows double period oscillations at short time scales. Their analysis, akin to that recently developed for MnSi [A. Amato {\it et al.}, Phys. Rev. B \textbf{89}, 184425 (2014)], provides an estimation of the field distribution induced by the Mn helical order at the muon site. The refined muon position agrees nicely with \emph{ab initio} calculations. With increasing temperature, an inhomogeneous fluctuating chiral phase sets in, characterized by two well separated frequency ranges which coexist in the sample. Rapid and slow fluctuations, respectively associated with short range and long range ordered helices, coexist in a large temperature range below \TN =170\,K. We discuss the results with respect to MnSi, taking the short helical period, metastable quenched state  and peculiar band structure of MnGe into account. 
\end{abstract}


\maketitle

\section{Introduction}
Spin fluctuations in itinerant systems have attracted strong attention since the pioneering work of Moriya\cite{Moriya1973}, providing a unified theory for the Curie-Weiss dependence of the spin susceptibility. Such fluctuations, either thermal or quantum, are usually precursor of a transition towards a magnetically ordered ground state. In critical phenomena, low energy fluctuations of the order parameter, extending over increasing length scales, yield a second order transition obeying Ginsburg-Landau universality laws. According to the Brazovskii scenario \cite{Brazovskii1975}, such a second order phase transition could be avoided if the fluctuations are strong enough, the system evading the associated entropy by undergoing a first order transition without any divergence of the correlation length. 
  
In B20 itinerant chiral magnets like MnSi, FeGe or MnGe, the crucial role of the spin fluctuations appears already in the ground state, when applied pressure induces a first order quantum transition from the ordered helical state to a non Fermi liquid state with partial magnetic order, involving chiral fluctuations of local magnetic moments \cite{Pfleiderer2004,Pedrazzini2007,Pappas2009,Hamann2011}. At finite temperature and under magnetic field, an intermediate chiral phase (also called fluctuation disordered regime) is stabilized between the ordered and paramagnetic phases. The nature of this phase has been discussed in the literature. One explanation invokes chiral mesophases in analogy with chiral nematics, including disordered of liquid phases composed of skyrmionic double twisted or multiply twisted spin textures\cite{Roessler2006,Pappas2009,Hamann2011,Wilhelm2012}. Another scenario assumes fluctuating helices \cite{Muhlbauer2009,Grigoriev2010} with finite lengthscale, isotropically distributed in space, and a fluctuation induced first order transition \cite{Janoschek2013} as explained by Brazovskii universality \cite{Brazovskii1975}. The chiral fluctuations could be the source of soft modes, stabilizing a skyrmion lattice phase in MnSi and FeGe just below the ordering transition\cite{Muhlbauer2009,Yu2010,Pedrazzini2007}.
 
In this series, MnGe stands out as a highly topical magnet, still poorly understood. Synthesized under high pressure and temperature \cite{Tsvyashchenko1984}, MnGe exists in metastable and powdered state only. The strong exchange interaction yields a high transition temperature (\TN\,= 170\,K) and ordered Mn moment $m_{\rm 0} = m_{\rm ord} (T \rightarrow 0) =$ 1.8(1) \mub\cite{Deutsch2014}, whereas the strong spin orbit coupling results in the shortest helix pitch (29 \,\AA \, at low temperature) of the B20 series \cite{Makarova2012}. Giant topological Hall effect (THE) and Nernst effect \cite{Kanazawa2011,Shiomi2013} make MnGe  promising for spintronic applications. In bulk MnGe in zero field, an helical multi-domain ground state was inferred from magnetic neutron diffraction, although a  more complex ground state involving a cubic lattice of skyrmions and anti-skyrmions was also proposed to account for the THE \cite{Kanazawa2012}. With increasing temperature, a fluctuating inhomogeneous chiral phase settles in, extending over an exceptionally broad temperature range \TN\ $\pm$ 70\,K. Fluctuations below \TN\ are a unique feature in the B20 series where they usually extend over a few degrees above \TN. 

In MnGe, these fluctuations, revealed by a broad susceptibility peak versus temperature, were studied by neutron diffraction and \Moss\ spectroscopy \cite{Deutsch2014b,Altynbaev2014}. These measurements suggest a qualitative picture of the chiral inhomogeneties.  Below \TN, long range ordered (LRO) helices coexist with short range ordered (SRO) fluctuating ones. Above \TN\, the static LRO helices disappear but SRO helices remain. Ferromagnetic correlations persist up to about 250\,K, with a coherence length below the typical helical wavelength. Low field magnetic irreversibilities are seen  even above, up to 300\,K at least, showing that some sort of slow dynamics coexists with rapid spin fluctuations deeply in the paramagnetic regime.

The nature and origin of the spin fluctuations in MnGe and the intrinsic inhomogenetity of its chiral order are matter of debate. The peculiar band structure of MnGe yields three possible states for the Mn moment, namely High Spin, Low Spin and Zero Spin (called HS, LS and ZS respectively)\cite{Roessler2012}, which can be stabilized depending on the interatomic distance. Therefore, the transition between spin states can be driven by an applied pressure. High pressure neutron diffraction \cite{Deutsch2014} shows that the collapse of the ordered Mn moment in the ground state occurs in two steps, around 7 GPa and  above  13  GPa respectively. X ray data measured up to 30 GPa \cite{Martin2016} suggest that the same scenario is at play for the local moment in the paramagnetic phase at 300\,K. 
 Altogether, the pressure data suggest a first order transition line between HS and LS states, stabilized  in a very large T range. This scenario yields a possible route for unconventional invar-like spin fluctuations, needed to accommodate HS and LS regions with different specific volumes.
 
We have investigated MnGe by $\mu$SR. At low temperature (T = 10\,K) when the frozen helical order is stabilized, we observe a complex oscillating asymmetry as in MnSi \cite{Amato2014}. Following Ref. \onlinecite{Amato2014}, we account for it by calculating the distribution of dipolar fields at the muon sites. Our analysis allows us to identify the muon stopping site in good agreement with an {\it ab initio} model and to determine the contact field on the muon site. With increasing temperature we use $\mu$SR to probe the spin fluctuations at a longer time scale (10$^{-6}$\,s) than the M\"ossbauer (10$^{-8}$\,s) or neutron (10$^{-11}$ to 10$^{-12}$\,s) probes. We deduce from our results the fluctuating fraction and relaxation rate versus temperature, describing the spin dynamics of the chiral fluctuations over a broad temperature range (10-300\,K) and time window.  The whole results are discussed with respect to the model MnSi case.
 
\section{Experimental details}
Polycrystalline MnGe was synthesized under 8 GPa in a toroidal high-pressure apparatus by melting reaction with Mn and Ge. The purity of the constituents was 99.9\% and 99.999\% for Mn and Ge respectively. The pellets of well-mixed powdered constituents were placed in rock-salt pipe ampoules and then directly electrically heated to T $\simeq$ 1600$^\circ$C. Then the sample was quenched to room temperature before releasing the applied pressure \cite{Tsvyashchenko1984}. The sample was the same as for the neutron experiments of Ref. \onlinecite{Deutsch2014,Deutsch2014b}. For the purpose of the $\mu$SR experiment it was sintered in a pellet of 13 mm diameter and 2 mm thickness, wrapped in a thin Al foil and placed in a silver sample holder. The $\mu$SR experiments were performed  on the GPS instrument at the Paul Scherrer Institut (PSI, Villigen, Switzerland), in the temperature range 10\,K-300\,K. In order too study the spin fluctuations we used longitudinal field $\mu$SR (LF-$\mu$SR) in a small field of 20 G to decouple the contribution of nuclear dipolar fields \cite{Hayano1979}. Measurements at selected temperatures in the range $1.5 \leq$ T $\leq 115$\,K were performed with a shorter time window (5 $\mu$s) and high statistics to extract the polarization oscillations induced by the helical order. Transverse field measurements (TF-$\mu$SR) were performed above 170\,K and the frequency shift was compared with the bulk magnetization to evaluate the contact field at the muon site in the paramagnetic region. To measure the magnetization, we used the same sample batch, magnetic field value (0.4 T) and cooling procedure as for the TF-$\mu$SR.    
\section{Results and analysis}
\label{sec:results}
\subsection{Helical order at low temperature}
\label{subsec:ordered_phase}

\begin{figure}
\begin{center}
\includegraphics[width=8cm]{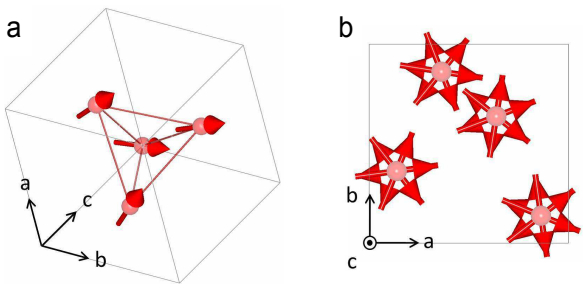}
\caption{Magnetic structure of MnGe deduced from neutron powder diffraction as seen along a) [1,1,$\overline{1}$] and b) [0,0,1] crystallographic directions. The latter picture consists in 6 consecutive unit cells, stacked along the $c$-axis.}
\label{fig:mag_struct}
\end{center}
\end{figure}

\begin{figure}
\begin{center}
\begin{tabular}{c}
\includegraphics[width=8cm]{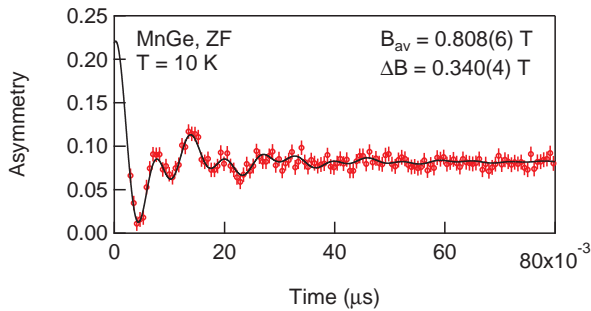}
\end{tabular}
\caption{Time dependence of the asymmetry measured at short times at T = 10\,K. 
 The solid curve is a fit of Eq. \ref{M1} to the data, see text.}
\label{asyshort}
\end{center}
\end{figure}

The helical magnetic structure of MnGe is shown in Fig \ref{fig:mag_struct}. 
The asymmetry of the positron emission, reflecting the time dependence of the muon polarization, recorded at low temperature (10\,K) with high statistics clearly exhibits an oscillating behaviour with double frequency (Fig. \ref{asyshort}). 
This oscillatory part of the muon polarization can be associated to the precession of the muon in the field distribution $D(B_{\text{loc}})$ induced by the helical order through
\begin{equation}
	P (t)= \int_{B_{\text{loc}}^{\text{min}}}^{B_{\text{loc}}^{\text{max}}} D(B_{\text{loc}}) \, \left[ \frac{1}{3} + \frac{2}{3} \cdot \cos \left(\gamma_{\mu} B_{\rm loc} \, t \right) \right] \, dB_{\text{loc}} \quad ,
\label{M0}
\end{equation} 
where $B_{\rm loc}^{\rm min}$ ($B_{\rm loc}^{\rm max}$) is the minimum (maximum) cutoff field value (see below) and $\gamma_{\mu} = 2\pi \cdot 135.5$ MHz.T$^{-1}$ the muon gyromagnetic ratio.

\emph{In a first step}, $P(t)$ was fitted by the following analytical expression, similar to that discussed in Eq. 17 of Ref.~\onlinecite{Amato2014}, which catches the essential features of the field distribution sensed by the muon

\begin{eqnarray}
   \nonumber
   P(t) &=&\frac{A(t)\, -b}{A_\text{0}-b}\\
   \nonumber 
	 &=& \frac{2}{3} \, J_{\text{0}}\left(\gamma_{\mu} \, \Delta B \, t\right) \, \cos{\left(\gamma_{\mu} B_{\text{av}} \, t +\psi\right)} \, e^{-\lambda_{\text{a}} \, t}\\
   &+& \frac{1}{3}e^{-\lambda_{\text{b}} \, t} \quad ,
\label{M1}
\end{eqnarray} 
  where $A_\text{0}$ = $A(t \rightarrow 0)$ is the effective initial asymmetry, $b$=0.007 is a temperature-independent background, $J_\text{0}$ a Bessel function of the first kind and $\psi$ a phase term. $B_{\text{av}}$ and $\Delta B$ are respectively the average field and width of the field distribution at the muon site. 

The two terms of the sum stem from the powder nature of the sample. Assuming random orientation of the helical domains and of the corresponding fields at the muon sites, in average 2/3 of the implanted muons precesses around a field perpendicular to their spin, whereas 1/3 of them experience a field along the initial muon polarization and do not precess. The relaxation rates $\lambda_{\text{a}}$ and $\lambda_{\text{b}}$ reflect as usual static or dynamic effects, which will be discussed in details in the next section by considering the full time scale. As in MnSi, the muon is assumed to stop at the Wyckoff position 4a of the space group P$2_{1}3$: I, located on the threefold rotation axis and the equivalent II, III and IV linked by this rotation (see Tab. \ref{tab:musites}). This will be justified below by \textit{ab initio} calculations. According to neutron diffraction data for MnGe, the helical order propagates along $[001]$ (or equivalent) direction, and not $[111]$ as found in MnSi. Therefore, in contrast to MnSi where the muon site 4a-I experiences a narrower field distribution than the others, the four muon sites are magnetically equivalent in MnGe, each of them feeling basically the same local field distribution.
 
The width of the field distribution $\Delta$B stems from the variation of the Mn moment direction from one cell to another induced by the helical order. A distance equivalent to the helical wavelength $\lambda_{\text{H}}$ = 2\ $\pi$/$|\mathbf{k}|$ where $\mathbf{k}$ is the wavevector of the helix is necessary to recover the same local magnetic environment at a muon site. It corresponds to about 6 unit cells for MnGe and 40 unit cells for MnSi.

\begin{table*}[!ht]
\caption{\label{tab:musites}Coordinates of the muon and its nearest Mn neighbors in the cubic unit cell.}
\begin{ruledtabular}
\begin{tabular}{cccccc}
	$4a$-site & & I & II & III & IV\\
	\hline
	Muon position & & (\xmu,\xmu,\xmu) & ($\frac{1}{2}$-\xmu,\xmub,$\frac{1}{2}$+\xmu) & ($\frac{1}{2}$+\xmu,$\frac{1}{2}$-\xmu,\xmub) & (\xmub,$\frac{1}{2}$+\xmu,$\frac{1}{2}$-\xmu)\\
	\hline
	& 1 & ($\frac{1}{2}$-$x$,1-$x$,$\frac{1}{2}$+$x$) & ($x$,$x$-1,$x$+1) & ($x$+1,$x$,$x$-1) & ($x$-1,$x$+1,$x$)\\
	Mn nearest neigbhors & 2 & ($\frac{1}{2}$+$x$,$\frac{1}{2}$-$x$,1-$x$) & ($x$-$\frac{1}{2}$,$\overline{x}$-$\frac{1}{2}$,1-$x$) & ($\frac{3}{2}$-$x$,$\overline{x}$,$x$-$\frac{1}{2}$) & ($\overline{x}$-$\frac{1}{2}$,1-$x$,$x$-$\frac{1}{2}$) \\
	& 3 & (1-$x$,$\frac{1}{2}$+$x$,$\frac{1}{2}$-$x$) & ($\overline{x}$,$x$-$\frac{1}{2}$,$\frac{3}{2}$-$x$) & (1-$x$,$x$-$\frac{1}{2}$,$\overline{x}$-$\frac{1}{2}$) & ($x$-$\frac{1}{2}$,$\frac{3}{2}$-$x$,$\overline{x}$)\\
\end{tabular}
\end{ruledtabular}
\end{table*}

\begin{figure}
\begin{center}
\begin{tabular}{c}
\includegraphics[width=8cm]{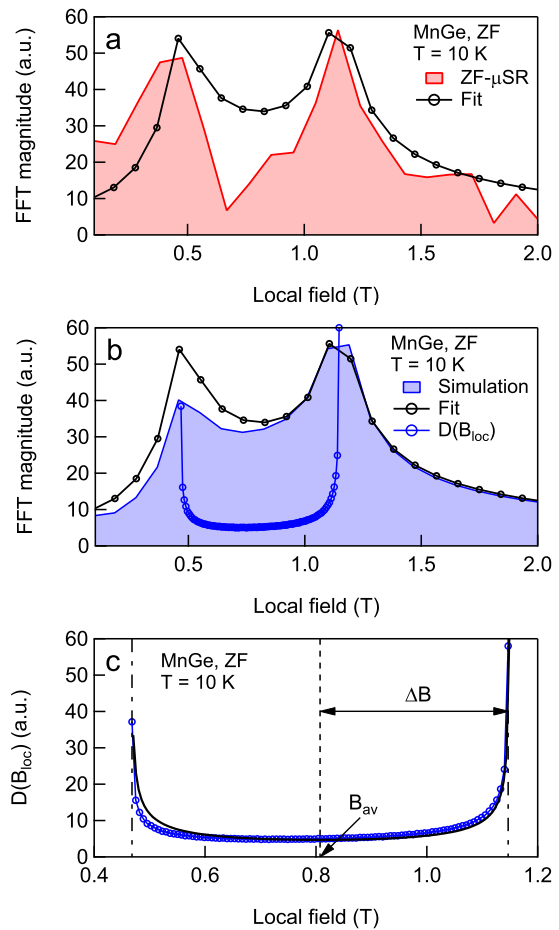}
\end{tabular}
\caption{ZF-$\mu$SR results for MnGe at 10 K. a) Fast Fourier Transforms (FFT) of the raw data (red area) and of the fitted curve (black symbols) (see Fig. \ref{asyshort}). b) FFT of the fitted curve now compared with the FFT of the simulated signal (blue area) and the calculated local field distribution $D(B_{\text{loc}})$ (blue symbols). c) Calculated local field distribution $D(B_{\text{loc}})$ (blue symbols). Black line is a fit of a shifted Overhauser function (Eq. \ref{M2}) to the data. See text for more details.}
\label{distrib}
\end{center}
\end{figure}

As shown in Fig. \ref{asyshort}, Eq. \ref{M1} yields a good fit of the experimental asymmetry. The fast Fourrier transform (FFT) of the fitted curve, plotted in Fig. \ref{distrib}, shows the experimental distribution of internal fields deduced from this analysis.  Following Ref. \onlinecite{Amato2014}, this distribution can be approximated by a a shifted Overhauser function  
  \begin{equation}
 D(B_{\text{loc}})=\frac{1}{\pi} \frac{1}{\sqrt{\Delta B^2-(B_{\text{loc}}-B_{\text{av}})^2}} \quad , 
\label{M2}
\end{equation} 
where $B_{\text{av}}=(B_{\text{max}}+B_{\text{min}})/2$ and  $\Delta B=(B_{\text{max}}-B_{\text{min}})/2$, with $B_{\text{min}}$ and $B_{\text{max}}$ the respective minimum and maximum cutoff values of the local field distribution. The peak width of the experimental field distribution arises from the limited time window of the FFT and the disorder inherent to the magnetic structure.   

\emph{In a second step},  the parameters of the field distribution were refined by performing a numerical calculation of the local field  $\mathbf{B}_{\text{loc}}$ at the muon site. The method we used for determining $\mathbf{B}_{\text{loc}}$ in MnGe is explained in full details in Appendix \ref{sec:appendixA}. For the sake of simplicity, we reproduce here the main steps only. 

For a given $\mathbf{R_{\mu}}$ vector joining the muon site to a Mn ion, the total field is defined as
    \begin{equation}
\mathbf{B}_{\text{loc}} \left(\mathbf{R_{\mu}}\right) = \mathbf{B}_{\text{dip}} \left(\mathbf{R_{\mu}}\right) + \mathbf{B}_{\text{cont}} \left(\mathbf{R_{\mu}}\right) \quad ,
\label{M3}
\end{equation} 
 where $\mathbf{B}_{\text{dip}}$ and $\mathbf{B}_{\text{cont}}$ are the \emph{dipolar} and \emph{contact} field respectively. 
  The dipolar field can be expressed as 
  \begin{equation}
\mathbf{B}_{\text{dip}} \left(\delta\right) =  \mathbf{C}_{\text{dip}} \cdot \cos \delta + \mathbf{S}_{\text{dip}} \cdot \sin \delta \quad ,
\label{M4}
\end{equation}
 where the lattice sums $\mathbf{C}_{\text{dip}}$  and $\mathbf{S}_{\text{dip}}$ are performed over a sphere of radius greatly overcoming the  helical wavelength (involving typically 10$^6$ unit cells) and $\delta$ =2$\pi \mathbf{k} \cdot \mathbf{R_{\mu}}$ can take all values between 0 and 2$\pi$ for an incommensurate structure, simulating all possible local environments along the spin helices for the muon. The contact field $\mathbf{B}_{\text{cont}}$ can be expressed as 
 \begin{equation}
\mathbf{B}_{\text{cont}} = \frac{A_{\text{cont}}}{N} \cdot \sum_{i=1}^{N}\mathbf{m_i} \quad , 
\label{M5}
\end{equation}
where  $A_{\text{cont}}$ is the contact coupling constant, $N = 3$ the number of Mn ions nearest neighbors of the muon and $\mathbf{m_i}$ their magnetic moment

\begin{equation}
	\mathbf{m_i} = m_{\rm ord} \cdot \left(\cos\varphi \cdot \mathbf{a} - \sin\varphi \cdot \mathbf{b}\right) \quad ,
	\label{M12}	
\end{equation}

where $m_{\rm ord}$ is the ordered moment, $\mathbf{a}$ and $\mathbf{b}$ being unit base vectors of the cubic unit cell and the minus sign accounting for the left handedness of the magnetic spirals of MnGe\cite{Grigoriev2013}. The phase term $\varphi=2\pi \mathbf{k} \cdot \mathbf{R_{\rm ij}}$, where $\mathbf{R_{\rm ij}}$ is a Mn-Mn vector, allows calculating the relative orientation of the Mn moments in the $\left(\mathbf{a},\mathbf{b}\right)$-plane, perpendicular to the helical wavevector $\mathbf{k}$. Such an approach yields large canting angle between neighboring Mn ions in MnGe ($\simeq 30^{\circ}$ between site 4a-I and 4a-II), as expected by the strong spin orbit coupling, while this angle is extremely small in MnSi ($\simeq 2^{\circ}$). Weak antiferromagnetic modes induced the Dzyaloshinskii-Moriya interaction, considered theoretically in Refs. \onlinecite{Chizhikov2012,Dmitrienko2012}, could lead to small out-of-plane tilts of the magnetic moments. They have not been detected yet by neutron diffraction and their existence would not change the conclusions of this paper. 

The harmonic approximation of Eq. \ref{M12} leads to the same dependence for the \emph{dipolar} and \emph{contact} contributions at the muon site, namely
\begin{equation}
\mathbf{B}_{\text{cont}} =  \mathbf{C}_{\text{cont}} \cdot \cos \delta + \mathbf{S}_{\text{cont}} \cdot \sin \delta \quad .
\label{M7}
\end{equation} 

Inserting Eq. \ref{M12} in Eq. \ref{M5} allows calculating the vectorial sums $\mathbf{C}_{\text{cont}}$ and $\mathbf{S}_{\text{cont}}$, performed over three Mn near neighbors of the muon site (their coordinates are given in Tab. \ref{tab:musites}).

\begin{figure}
\begin{center}
\begin{tabular}{c}
\includegraphics[width=8cm]{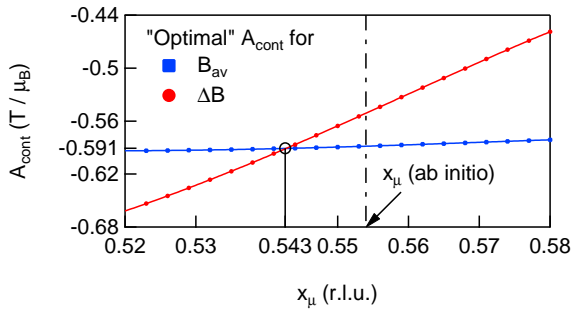}
\end{tabular}
\caption{Determination of the couple of optimized parameters (\xmu, $A_{\text{cont}}$): the muon fractional coordinate  \xmu\,= 0.543 and and the contact coupling constant $A_{\text{cont}}$ = -0.591 T.$\mu_{\rm B}^{-1}$ optimize the calculated distribution of internal field as compared with the one measured at 10\,K. See text.}
\label{optimiz}
\end{center}
\end{figure}   

Experimentally, the distribution of local fields at the muon site $D(B_{\rm loc})$ can be accessed through the real part of a Fast Fourier Transform (FFT) of the early time zero field $\mu$SR (ZF-$\mu$SR) signal (see Eq. \ref{M0}). The result obtained at 10 K is shown in Fig. \ref{distrib}a and compared with the FFT of the fit of Eq. \ref{M2} to the data.

In order to check the origin of the derived spectrum, mainly composed of two maxima, we have computed the expected $D(B_{\rm loc})$ in MnGe by setting the magnetic and crystal parameters from the neutron data\cite{Makarova2012,Deutsch2014}  measured on the same sample at the same temperature ($T = 10$\, K), namely the lattice constant $a = 4.769$ \AA, the fractional coordinate of the Mn ions $x = 0.138$ r.l.u., the helical wavelength $\lambda_{\text{H}} = 28.7$ \AA~and the ordered Mn magnetic moment $m_{\rm ord} = 1.83$ \mub. For these values, the local field distribution is calculated by sampling $10^4$ values of $\delta$, and optimized by a stepwise variation of the muon site coordinate \xmu~and the value of the contact coupling constant $A_{\text{cont}}$. The {\it contact} contribution to the total field is found to be about twice as large as the {\it dipolar} one and it has opposite sign. 

In Fig. \ref{distrib}b, we plot the calculated $D(B_{\rm loc})$ which is in excellent agreement with the experimental one. In addition, we display a simulation of the expected ZF-$\mu$SR signal, generated by using the calculated $D(B_{\rm loc})$ and considering the limited time frame, as well as the observed fast damping of the measured asymmetry induced by the disorder inherent to the magnetic structure ($\lambda_{\text{a}}$ term in Eq. \ref{M1}). These combined effects lead to a broadening of the Fourier transform, nicely reproducing the experimental spectrum (Fig. \ref{distrib}b).

In summary, we find the same field distribution for the 4 muon sites, which is quite close to the Overhauser function (Fig. \ref{distrib}c). The agreement between the calculated distribution and the experimental one deduced by FFT of the $\mu$SR signal is very sensitive to the position of the muon site and the value of the coupling constant.  We find only $\it{one}$ couple of parameters optimizing both the width and average value for the field distribution (Fig. \ref{optimiz}), with values \xmu\,= 0.543 and $A_{\text{cont}}$ = -0.591 T.$\mu_{\rm B}^{-1}$.  The average field and field distribution are respectively B= 0.808(6)\,T and $\Delta B$= 0.340(5)\,T. 

\begin{figure}
\begin{center}
\begin{tabular}{c}
\includegraphics[width=8cm]{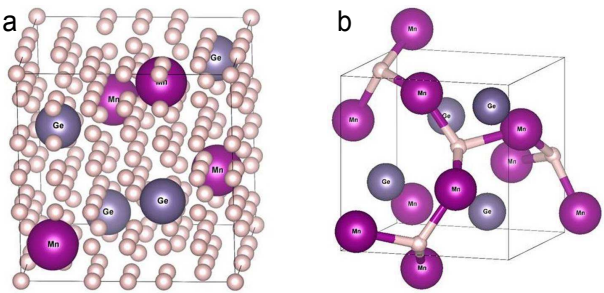}
\end{tabular}
\caption{Result of the \textit{ab initio} calculation. a) Interstitial positions candidates for a muon site. b) Final position resulting from the structural relaxation, showing the three Mn neighbors of each muon site.}
\label{ab_initio}
\end{center}
\end{figure}   
The position of the muon site was estimated independently by \textit{ab initio} calculation as for MnSi \cite{Bonfa2015}. Details of the calculations for MnGe are given in Appendix B. The $6 \times 6 \times 6$ grid used to sample the interstitial space associated with the unit cell volume (Fig. \ref{ab_initio}a)  reduces to only $\it{one}$ interstitial position that corresponds to a candidate muon site having fractional coordinates (\xmu,\xmu,\xmu) with \xmu= 0.554 in the unit cell. This value compares well with the experimental value \xmu = 0.543 deduced from the analysis of our ZF-$\mu$SR data. The output of the structural relaxation showing the muon site with respect to Mn and Ge ones is shown in Fig. \ref{ab_initio}b.
\begin{figure}
\begin{center}
\begin{tabular}{c}
\includegraphics[width=8cm]{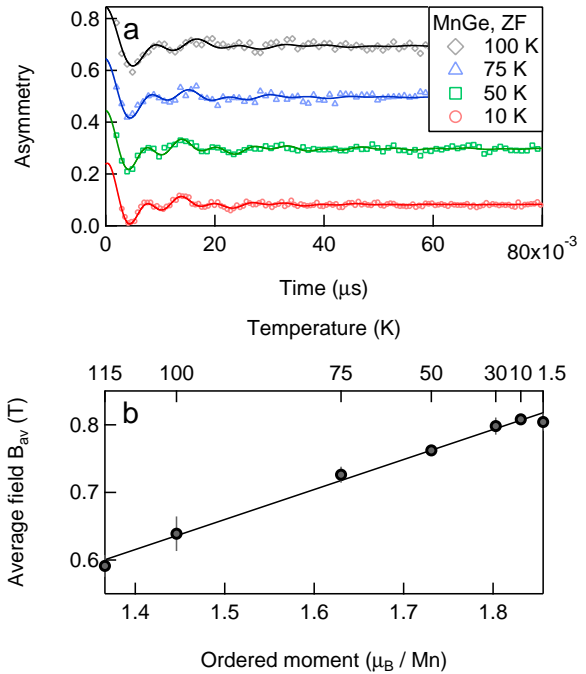}
\end{tabular}
\caption{a) Asymmetry measured in zero field and at short times for several temperatures. b) Average internal field B$_{\rm av}$ compared with the ordered Mn moment measured by neutron diffraction.}
\label{asyshort_T}
\end{center}
\end{figure}   
Asymmetry patterns, recorded at selected temperatures, have been analyzed with the same procedure (example spectra are displayed in Fig. \ref{asyshort_T}a). With increasing temperature, the ordered magnetic moment strongly decreases whereas the propagation vector increases, and these temperature variations are known precisely from neutron diffraction \cite{Deutsch2014,Deutsch2014b}. The average field on the muon site is proportional to the ordered Mn moment (Fig. \ref{asyshort_T}b), which confirms the validity of the analysis.  
\begin{figure}
\begin{center}
\begin{tabular}{c}
\includegraphics[width=8cm]{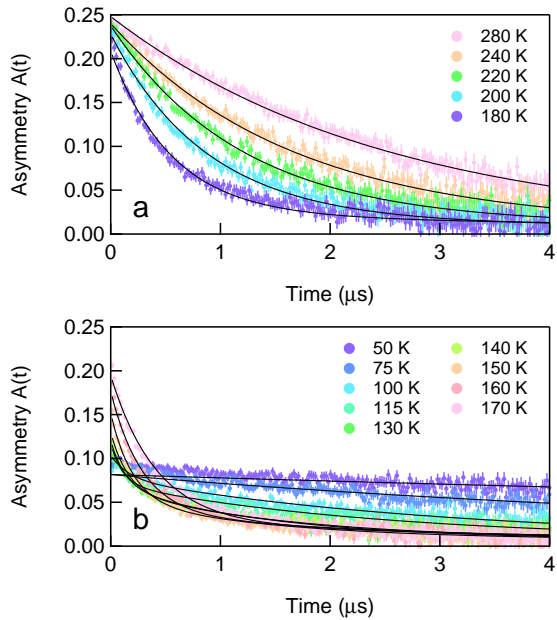}
\end{tabular}
\caption{$\mu$SR spectra recorded in a 20 G longitudinal field and at selected temperatures a) above and b) below \TN. Solid curves are fits with Eq. \ref{M8}, as described in text.}
\label{asy}
\end{center}
\end{figure}   

\subsection{Spin fluctuations and phase separation}
With increasing temperature, the decrease of the ordered helical moment is associated with the onset of strong spin fluctuations, as MnGe enters the inhomogeneous fluctuating chiral phase. The time dependence of the asymmetry, recorded in a 20\,G longitudinal field, is shown in Fig. \ref{asy} (a, b) for temperatures above and below \TN\ respectively. At 10\,K, a fast depolarization occurs due to the helical order, so that the effective initial asymmetry $A_{\rm 0} = A(t \rightarrow 0)$ falls down to one third of the total asymmetry $A_{\text{tot}} = 0.245$, measured well above \TN. Upon heating, $A_{\rm 0}$ starts increasing around 100\,K, when a fluctuating paramagnetic fraction starts coexisting with the ordered one. 

\begin{figure}
\begin{center}
\begin{tabular}{c}
\includegraphics[width=8cm]{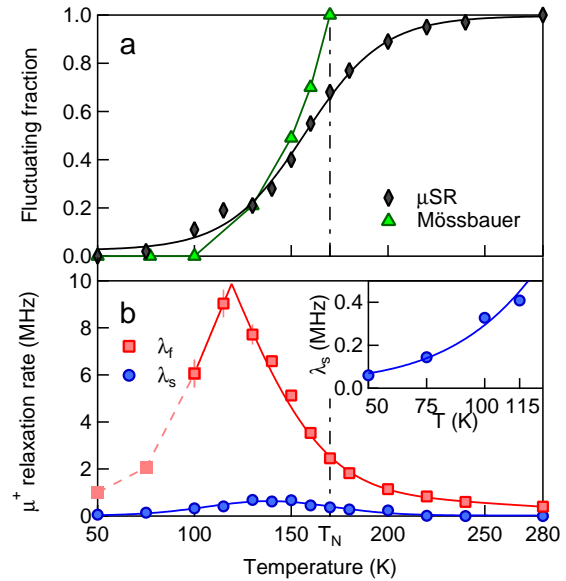}
\end{tabular}
\caption{a) Fluctuating phase fraction derived from a fit of Eq. \ref{M8} to the $\mu$SR spectra, in comparison with the paramagnetic fraction deduced from \Moss\ spectroscopy on a $^{57}$Fe-doped MnGe sample\cite{Deutsch2014b}. b) Fast and slow relaxation rates obtained through the same analysis. Inset: slow fluctuations are observed at a small but finite rate $\lambda_{\rm s}$ down to the lowest temperatures. In all panels, lines are guides to the eye.} 
\label{lambda_ratio}
\end{center}
\end{figure}

In order to account for the phase separation between frozen (LRO) helices and fluctuating (SRO) ones \cite{Deutsch2014b}, the following functional form was fitted to the long time tail of the $\mu$SR spectra (t $>$ 0.1 $\mu$s) at each temperature
 \begin{equation}
A(t)= \left(A_{\text{tot}}-b\right)\cdot\left[\frac{1-f}{3} e^{-\lambda_{\text{s}} t} +f e^{-\lambda_{\text{f}} t}\right]+b \quad ,
\label{M8}
\end{equation}
  where $1-f$ and $f$ are the volume fractions of the long range ordered  and short range ordered phases and $\lambda_s$ and $\lambda_f$  the corresponding relaxation rates, associated with $\textit{slow}$ and $\textit{fast}$ relaxations respectively. $\it{b}$ is a small temperature independent background arising from the muons falling into the sample holder, which was measured and subsequently fixed to $7 \cdot 10^{-3}$. In Eq. \ref{M8}, one neglects the oscillations at short times ($t < 0.1~\mu$s) discussed in Sec. \ref{subsec:ordered_phase}. As shown in Fig. \ref{asy}, good fits were obtained in the whole temperature range. The temperature dependence of the volume fraction $f$ and relaxation rates $\lambda_{\rm s}$ and $\lambda_{\rm f}$ is displayed in Fig. \ref{lambda_ratio}. The volume fraction associated with the fast relaxation is in good qualitative agreement with the paramagnetic fraction deduced from published \Moss\ data\cite{Deutsch2014b}. Below 100 K, the latter becomes very small ($f \rightarrow 0$) and the associated relaxation rate $\lambda_{\rm f}$ becomes less reliable. However, the persistence of magnetic fluctuations down to the lowest temperatures is shown by the observation of a finite value for $\lambda_{\rm s}$ (see inset of Fig. \ref{lambda_ratio}).  

\begin{figure}
\begin{center}
\begin{tabular}{c}
\includegraphics[width=8cm]{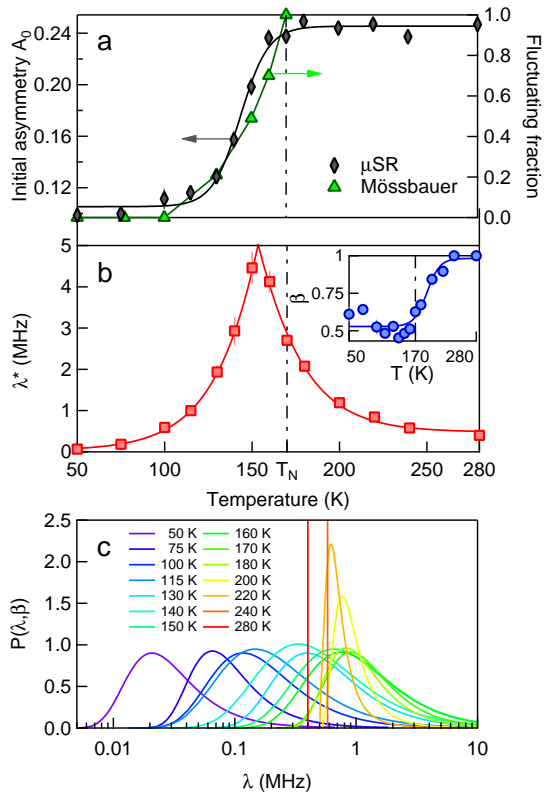}
\end{tabular}
\caption{a) Effective initial asymmetry of the $\mu$SR signal as a function of temperature. The sharp increase of $A_{\rm 0}$ upon heating reflects the melting of the long-range helimagnetic order. b) Characteristic relaxation rate $\lambda^{*}$ obtained from a fit of Eq. \ref{M13} to the data (main panel), along with the stretching exponent $\beta$ (inset). In panels a and b, lines are guide to the eye. c) Calculated relaxation rates distribution computed with the help of experimentally found values for $\lambda^{*}$ and $\beta$ (see text).} 
\label{lambda_distribs}
\end{center}
\end{figure}

Instead of a discrete relaxation spectrum composed of two well-defined rates, we alternatively considered a broad distribution of independent relaxation channels, as considered for instance in spin glasses\cite{Campbell1994,Keren1996,Pickup2009}, superparamagnetic particles\cite{Bewley1998,Jackson2000} or some frustrated pyrochlore magnets\cite{Bert2006,Dalmas2006,Guo2014}. In such case the asymmetry can be described by a continuous sum of exponential decays, usually modeled as a stretched exponential 
	
\begin{equation}
A(t)= \left(A_{\rm 0}-b\right)\cdot e^{-\left(\lambda^{*} t\right)^{\beta}} + b \quad ,
\label{M13}
\end{equation}

where $\lambda^{*}$ is the characteristic muon relaxation rate, $\beta$ a stretching exponent and $b$ is defined as explained above. Note that the effective initial asymmetry $A_{\rm 0}$ is now refined, as opposed to $A_{\rm tot}$ which was fixed in Eq. \ref{M8}. This choice allows focusing on the spin dynamics of the system \emph{only} by filtering out the fast decay of the measured asymmetry induced by the magnetic order at low temperature. Fitting Eq. \ref{M13} to the data in the range $50 \leq T \leq 280~\text{K}$ yields an equally good statistical agreement as compared with the two fractions model. The parameters derived from this procedure are displayed in Fig. \ref{lambda_distribs}a and \ref{lambda_distribs}b. 

The thermal evolution of $A_{\rm 0}$ is known to be very sensitive to the onset of long-range magnetic order\cite{Dalmas1997}. Indeed, in MnGe, $A_{\rm 0}$ is found to saturate at a value $A_{\rm 0} (T > T_{\rm N}) = A_{\rm tot} = 0.245$ above $T_{\rm N}$ and to drop to about one third of this value in a 70 K interval below $T_{\rm N}$, in perfect agreement with neutron diffraction results of Ref. \cite{Deutsch2014b}. In order to describe the overall muon relaxation spectrum described by Eq. \ref{M13}, we follow the procedure detailed in Ref. \onlinecite{Johnston2006} where the probability distribution function $P\left(\lambda\right)$ is introduced and defined as 
\begin{equation}
\frac{1}{\lambda^{*}} \, \int_{0}^{\infty} P\left(\lambda,\beta\right) \, e^{-\lambda \, t} \, d\lambda = e^{-\left(\lambda^{*} t\right)^{\beta}} \quad . 
\label{M14}
\end{equation}
Since $0.5 \leq \beta \leq 1$ (see inset of Fig. \ref{lambda_distribs}b), $\lambda^{*}$ can be regarded as an accurate estimate of the median of $P\left(\lambda,\beta\right)$ in the whole temperature range. We thus retain this value as representative of the probed physics and will use it in our evaluation of the electronic spin fluctuation frequencies (Sec. \ref{sec:discuss}). 
For the sake of completeness, we display calculated $P\left(\lambda,\beta\right)$ in Fig. \ref{lambda_distribs}c, allowing to follow the evolution of the shape of the relaxation spectrum as a function of temperature. While broadened spectra are observed up to about 220 K, a Dirac $\delta$ function -corresponding to a single frequency spectrum- is recovered at higher temperatures, when magnetic correlations are becoming small with respect to the thermal energy.

As a partial conclusion, we stress that irrespective of the model used to describe the long time LF-$\mu$SR spectra, our data strongly indicates that spin fluctuations are surviving deep inside the magnetically ordered phase, at odds with other known cubic Dzyaloshinskii-Moriya helimagnets.
 
\subsection{Transverse field measurements in the paramagnetic regime}
\begin{figure}
\begin{center}
\begin{tabular}{c}
\includegraphics[width=8cm]{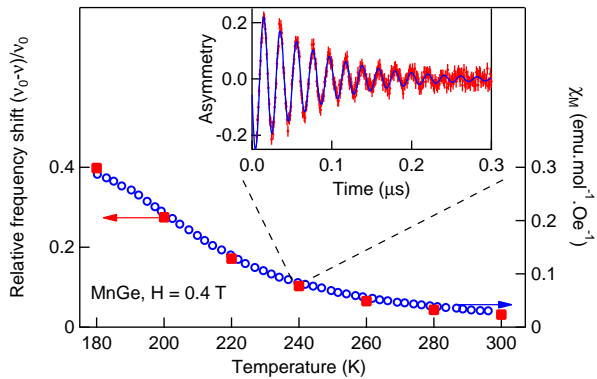}
\end{tabular}
\caption{Relative frequency shift $(\nu_0-\nu)/\nu_0$ (red squares, left scale) deduced from TF-$\mu$SR spectra measured in the paramagnetic regime, in comparison with the macroscopic susceptibility (blue circles, right scale). Both sets of data were measured at 0.4\,T. A typical TF-$\mu$SR spectrum taken at 240\,K is shown in inset.}
\label{TFMuSR}
\end{center}
\end{figure}

In a magnetic field H perpendicular to its initial momentum, the muon spin precesses with a frequency $\nu$ related to the susceptibility of the bulk material. A typical spectrum measured in MnGe in the paramagnetic regime with a field of 0.4\,T is shown in the inset of Fig. \ref{TFMuSR}. The frequency shift $\nu_0-\nu$ with respect to the frequency $\nu_{0}$ = 54.2\,MHz in the Ag sample holder can be expressed as
\begin{equation}
\nu_{0}-\nu=  \frac{\gamma_{\mu}}{2\pi} \cdot \langle B_{\text{loc}} \rangle \quad ,
\label{M9}
\end{equation}
where the local field
\begin{equation}
 B_{\text{loc}}=B_{\text{ext}}+B_{\text{cont}}+B_{\text{dip}}+B_{\text{Lor}}+B_{\text{dem}}
\label{M10}
\end{equation} 
involves terms corresponding to the applied, contact, dipolar, Lorentz and demagnetizing fields respectively, which must be averaged over all orientations. Due to this average, the contribution of the dipolar term cancels, and the contact term reduces to B$_{\text{cont}}$= $A_{\rm cont} \, \chi_{\rm M} \, H$, where $\chi_{\rm M}$ is the isotropic susceptibility of a MnGe mole of volume $V_{\rm M}$, and A$_{\text{cont}}$ is the isotropic average of the hyperfine contact tensor. The demagnetizing and Lorentz fields are respectively equal to $B_{\text{dem}} = -4\pi \, N \frac{\chi_{M}}{V_{\rm M}} \, H$ and $B_{\text{Lor}} = \frac{4\pi}{3} \, \frac{\chi_{\rm M}}{V_{\rm M}}H$.  
As shown in Fig. \ref{TFMuSR}, the temperature dependence of the $\mu$SR shift compares well with that of the macroscopic susceptibility. Assuming a demagnetization factor corresponding to the shape of the sintered pellet used in our experiment ($N \simeq 0.8$), we obtain a contact coupling constant $A_{\text{cont}} = -0.45(11) \text{~T.}\mu_{\rm B}^{-1}$ from the measured frequency and macroscopic susceptibility ($\nu$\,= 48.7 MHz  and $\chi_{\rm M}= 8.3 \cdot 10^{-2}$ emu.mol$^{-1}$.Oe$^{-1}$ at 240 K). On the other hand, if we consider a distribution of individual grain shapes within the sample, with $0 \leq N \leq 1$ in the extreme case, we end up with $A_{\text{cont}} = -0.6(2)$ T.$\mu_{\rm B}^{-1}$, a value in even closer agreement with our experimental determination by ZF-$\mu$SR (see Section \ref{subsec:ordered_phase}). 
\section{ Discussion}
\label{sec:discuss}

The origin of multiple frequency oscillations in $\mu$SR spectra is usually ascribed to magnetic unequivalent muon stopping sites, as it occurs for instance in Cu$_2$OSeO$_3$ helical magnet \cite{Maisuradze2011,Lancaster2015}. An alternative explanation previously proposed in MnSi considers the formation of a spin-polaron state, made by a localized electron state mediating ferromagnetic interactions with the neighbouring magnetic ions, and inducing two different states  for the muon \cite{Storchak2011}. Here we can exclude the presence of crystallographic unequivalent muon sites. The two frequency asymmetry observed in ZF-$\mu$SR is attributed to the field distribution on a given muon site induced by the helical order, and the approach of Ref. \onlinecite{Amato2014} in MnSi single crystal is supported in MnGe with polycrystalline form, shorter helical period, higher moment, and different local symmetry induced by the propagation vector. In MnGe, the magnetic environment of the Mn neighbors is also far from the ferromagnetic droplet required to localize an electron, because of the large canting between near neighbor Mn moments.
 
In contrast with MnSi where the site 4a I merely experiences a unique field, in MnGe  the four 4a Wickoff sites occupied by the muon feel the same field distribution. 
The fitted position of the muon site \xmu= 0.543 compares well with the value \xmu= 0.554 deduced from \textit{ab initio} model. At low temperature (T $\ll$ \TN) the average field and width of the field distribution in MnGe can be compared with the values relative to the sites II, III, and IV in MnSi (see Tab \ref{tab:comparison}). The average field is much larger in MnGe than in MnSi ($B_{\text{av}}^{\text{MnGe}}/B_{\text{av}}^{\text{MnSi}}$ $\sim$ 5), roughly reflecting the ratio of dipolar contributions $m_{\rm ord}/a^3 \sim 4$. On the other hand, the relative widths of the field distributions $\eta = \Delta B / B_{\text{av}}$ are only slightly different, namely $\eta_{\text{MnGe}} / \eta_{\text{MnSi}} \simeq 1.15$. Indeed, if we would expect measuring a broader local field distribution for a magnetic structure having a longer periodicity, we must also account for the  different ordered moment values. This implies a scaling of the form $\eta \propto 1/(m_{\rm ord}\cdot\lambda_{\text{H}})$ which, given the material-specific parameters in Tab. \ref{tab:comparison}, yields $\eta_{\text{MnGe}} / \eta_{\text{MnSi}} \simeq 1.37$ in agreement with the experimental value. 

\begin{table}[!ht]
\caption{\label{tab:comparison}Comparison of structural, magnetic and ZF-$\mu$SR data for MnGe and MnSi. In the case of MnSi, $B_{\text{av}}$ and $\Delta B$ are considered for site II only (see Ref. \onlinecite{Amato2014}).}
\begin{ruledtabular}
\begin{tabular}{lcc}
	& MnGe (10 K) & MnSi (5 K)\\
\hline
$T_{\text{N}}$ [K] & 170 & 29.5\\	
$a$ [\AA] & 4.769 & 4.558\\
$\lambda_{\text{H}}$ [\AA] & 28.7 & 180\\
$m_{\rm ord}$ [$\mu_{\text{B}}$/Mn] & 1.83 & 0.4\\
$B_{\text{av}} [T]$ & 0.808(6) & 0.152\\
$\Delta B$ [T] & 0.340(5) & 0.056\\
$\Delta B / B_{\text{av}}$ & 0.42 & 0.37 \\
$A_{\text{cont}}$ [T.$\mu_{\rm B}^{-1}$] & -0.591 & -0.518 \\
\xmu [r.l.u] & 0.543 & 0.532\\
\end{tabular}
\end{ruledtabular}
\end{table}
 
The dominant contribution to the local field measured by ZF-$\mu$SR is played by the contact term, which amounts to twice the contribution of the dipolar term. As for the contact coupling constants A$_{\text{cont}}$, we find very similar values in MnGe and MnSi ($\sim$ -0.55 T.$\mu_{\rm B}^{-1}$) from the analysis of the ZR-$\mu$SR spectra measured in the ordered state at low temperature. The value found for MnGe by TF-$\mu$SR in applied field in the paramagnetic state ($\sim -0.45 \text{~T.}\mu_{\rm B}^{-1}$) is slightly smaller, but entailed by a large error bar due to the uncertainty on the demagnetization factor for a powdered sample.

The time dependence of the asymmetry in the full time window reflects the spin fluctuations in the inhomogeneous chiral phase. Assuming a dynamical phase separation in MnGe, $\mu$SR probes two different electronic relaxation rates for the Mn moments, reflected in typical values of the muon relaxation rates differing by more than an order of magnitude. The phase ratio between the $\textit{slow}$ and $\textit{fast}$ fluctuating fractions can be compared with the ratio of \, $"$frozen$"$ to $"$paramagnetic$"$ fractions deduced from \Moss\ spectroscopy. The latter probes a shorter and narrower time window (10$^{-7}$s to 10$^{-9}$s) than $\mu$SR, but the qualitative agreement is very good. A more phenomonological description, assuming a broad distribution of frequencies with a shape evolving with temperature (as traced by the change of stretching exponent $\beta(T)$ in Fig. \ref{lambda_distribs}c), is also compatible with the $\mu$SR data, and it might explain why in the \Moss\ spectroscopy with narrow time window, one does not observe any relaxing behavior.
 
Altogether, fast and slow spin fluctuations can coexist over a temperature range extended by the muon probe to about 50-250 \,K, a huge range with respect to other chiral magnets. We relate the \emph{slow} dynamics to LRO helices and the \emph{fast} one to SRO helices or ferromagnetic correlations (namely incomplete helices), recalling that both have been observed, coexisting in the same temperature range, by neutron diffraction\cite{Deutsch2014b} and small-angle scattering\cite{Altynbaev2014}. We tentatively attribute the origin of this magnetic inhomogeneities in a chemically pure compound to the peculiar MnGe band structure and quenched state, inducing metastable low spin states in a dominant high spin state. 
 
\begin{figure}[!ht]
\begin{center}
\begin{tabular}{c}
\includegraphics[width=8cm]{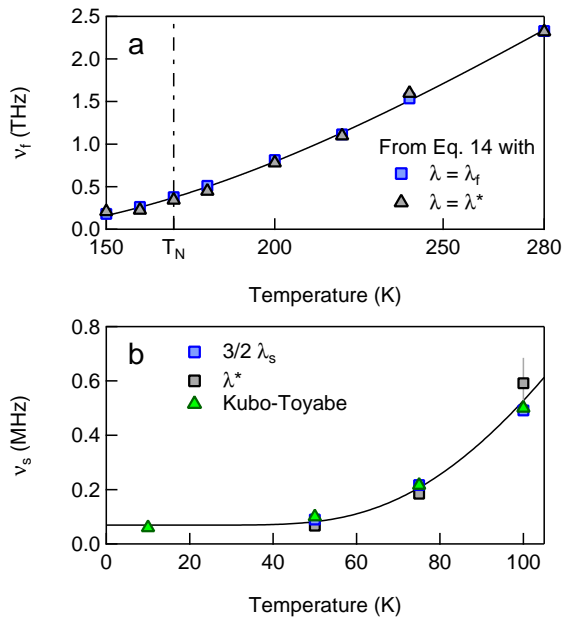}
\end{tabular}
\caption{a) Thermal evolution of the {\it fast} spin fluctuation frequency $\nu_{\rm f}$, derived from our LF-$\mu$SR measurements by virtue of Eq. \ref{M11}. b) {\it Slow} fluctuations frequency as a function of temperature obtained in the low temperature limit. Solid line are guides to the eye.}
\label{NuvsT}
\end{center}
\end{figure}

In the paramagnetic regime, in the limit of fast fluctuations, the muon relaxation rate $\lambda$ is related to the typical electronic spin fluctuation frequency $\nu_{\rm f}$ by
\begin{equation}
\lambda = \frac{2 \cdot \gamma_{\mu}^2 \cdot \langle B^2 \rangle }{\nu_{\rm f}} \quad ,
\label{M11}
\end{equation}
where $\langle B^2 \rangle$ is the second moment of the distribution of fluctuating field experienced by the muons\cite{Hayano1979}. In the simplest case,  the random fluctuations of the magnetic moments well above \TN\ yield an average field $\langle B \rangle = 0$ and a width of the field distribution $\langle B^{2} \rangle \simeq B_{\text{av}}^2$, where $B_{\text{av}} \simeq 0.8~\text{T}$ is the internal field measured at low temperature. This yields a typical frequency of the fluctuations $\nu_{\rm f}\sim$ 2.5 THz at 300\,K, a frequency range which could be probed {\it e.g.} by inelastic neutron scattering. With decreasing temperature, the increase of $\lambda_{\rm f}$ (Fig. \ref{lambda_ratio}b) or $\lambda^{*}$ (Fig. \ref{lambda_distribs}b) reflects the slowing down of the fluctuations when approaching the transition. The evolution of the fluctuation frequency derived from Eq. \ref{M11} is illustrated in Fig. \ref{NuvsT}. In the paramagnetic regime, $\nu_{\rm f}$ is a monotonously increasing function of temperature (Fig. \ref{NuvsT}a). Below $T_{\rm N}$, the $\mu$SR signal becomes quickly dominated by the growing ordered fraction within the sample. In order to get deeper insight into the critical dynamics of MnGe, one should use space-resolved techniques which allows determining the $Q$-dependence of the relaxation spectrum in contrast to {\it local} probes such as $\mu$SR or \Moss ~where all length scales contribute to the signal.  

Conversely, in the ordered regime at low temperature, the residual fluctuations probed by the muon in an almost static distribution of internal fields can be modeled  by the dynamical Kubo-Toyabe function which, in the slow hopping limit, extrapolates to\cite{Hayano1979} 

\begin{equation}
	A(t) \sim \frac{A_{\rm tot}}{3} \cdot \exp\left(-2/3 \, \nu_{\rm s} \, t\right) \quad ,
\label{M15}
\end{equation}
where $\nu_{\rm s}$ is the characteristic spin fluctuation frequency within the ordered phase.
Neglecting the oscillations at short times, the long time tail of the LF-$\mu$SR spectra indeed show a residual slope with respect to the 1/3-plateau. Estimating $\nu_{\rm s}$ via {\it (i)} a direct fit of Eq. \ref{M15} to the data, {\it (ii)} $\lambda_{\text{s}}~(=~3/2 \, \nu_{\rm s}$) in Eq. \ref{M8} or {\it (iii)} $\lambda^{*}$ in Eq. \ref{M13} offers a perfect correspondence, as illustrated in Fig. \ref{NuvsT}b.
%
These slow fluctuations, inaccessible to the \Moss\ and neutron probes, could correspond to thermally activated lock-in and unlock-in processes for the helices. 

\section{Conclusion}
The distribution of internal fields probed by $\mu$SR in MnGe is qualitatively similar to that in MnSi, and quantitatively explained by its peculiar helical order, with shorter helical pitch, higher Mn moment and different propagation vector. As in MnSi, we identify a unique Wyckoff site for the muon. Our analysis therefore supports the helical order as a direct origin of a double frequency time variation of the asymmetry for both compounds, without need to invoke either a spin polaron state or multiple muon sites. The main anomaly of MnGe lies in the inhomogeneous fluctuating chiral phase, which can be studied by the $\mu$SR technique. Considering a two-phase system or a broad relaxation spectrum yields, in both cases, spin dynamics in a large frequency range. \emph{Slow} and \emph{fast} spin fluctuations are found to coexist within a large temperature interval, extended by the muon probe to about 50-250\,K, in stark contrast with other chiral magnets where fluctuations are usually confined nearby \TN. Altogether, the range of fluctuations probed by the muon extends from the THz regime at 300\,K down to a tens of kHz at 10\,K. The exceptionally broad temperature range and time scale where different spin dynamics coexist may result from the metastable character and invar-like band structure of MnGe. 
 
\section*{Acknowledgements}

Part of this work was performed at the Swiss Muon Source (S$\mu$S, PSI, Villigen, Switzerland).
We are grateful to P. Dalmas de R\'eotier for very useful discussions.
The post doc training of N. Martin is funded by the LabEx Palm. Public grant from the "Laboratoire d'Excellence Physics Atom Light Mater" (LabEx PALM) overseen by the French National Research Agency (ANR) as part of the "Investissements d'Avenir" program (reference: ANR-10-LABX-0039). 
The post doc training of M. Deutsch was funded by the ANR (DYMAGE).
D. Andreica acknowledges partial financial support from Romanian UEFISCDI Project No. PN-II-ID-PCE-2011-3-0583 (85/2011).
P. Bonf\`a thanks the computing resources provided by CINECA within the Scientific Computational Project CINECA ISCRA Class C (Award HP10C5EHG5, 2015) and STFC's Scientific Computing Department.
L.N. Fomicheva and A.V. Tsvyashchenko acknowledge the support of the Russian Foundation for Basic Research (Grant 14-02-00001). 

\appendix
\section{Dipolar and contact field calculation}
\label{sec:appendixA}

We derive here the method used for computing the \emph{local field} at the muon site (adapted from Refs. \onlinecite{Andreica2001,Schenck2001,Schenck2003}).\\

The \emph{dipolar field} created by an assembly of magnetic moments $\mathbf{m_i}$ at the muon site can be calculated through:

\begin{equation}
	\mathbf{B_{\text{dip}}} (\mathbf{R_\mu}) = \frac{\mu_{\text{0}}}{4 \pi} \cdot \sum_{\text{i}} \left[\frac{3\mathbf{r_i}(\mathbf{m_i}\cdot\mathbf{r_i})}{\left|\mathbf{r_i}\right|^{5}}-\frac{\mathbf{m_i}}{\left|\mathbf{r_i}\right|^{3}}\right]
	\label{A1}	
\end{equation}

where $\mathbf{r_i} = \mathbf{R_i} - \mathbf{R_\mu}$ is the vector connecting the $i^{\text{th}}$ magnetic moment and the muon. For a helicoidal magnetic structure, $\mathbf{m_i}$ reads:

\begin{equation}
	\mathbf{m_i} = m_{\rm ord} \cdot \left(\cos\varphi_{\text{i}} \cdot \mathbf{a} \pm \sin\varphi_{\text{i}} \cdot \mathbf{b}\right)
	\label{A2}	
\end{equation}

where $m_{\rm ord}$ is the staggered moment, $\mathbf{a}$ and $\mathbf{b}$ being orthogonal unit vectors in the plane perpendicular to the propagation vector $\mathbf{k}$ of the spin helix. The sign $\pm$ allows for choosing between left- or right-handed spin spirals. In the cubic basis, $\mathbf{a} = (1,0,0)$, $\mathbf{b} = (0,1,0)$ and $\mathbf{k} = (0,0,\frac{2\pi\zeta}{a})$, with $\zeta$ the index of the (incommensurate) magnetic structure and $a$ the cubic lattice constant. The phase term $\varphi_{\text{i}}$ in Eq. \ref{A2} reads

\begin{equation}
	\varphi_{\text{i}} = \mathbf{k} \cdot (\mathbf{R_i} - \mathbf{R_{\text{ref}}}) = \mathbf{k} \cdot (\mathbf{r_i} + \mathbf{R_\mu} - \mathbf{R_{\text{ref}}})
	\label{A3}	
\end{equation}

where $\mathbf{R_{\text{ref}}}$ is the position of a reference magnetic ion ({\it i.e.} for which $\alpha = 0$). Thus, the cosine and sine terms in \ref{A2} can be rewritten owing to trigonometric identities:

\begin{widetext}
\begin{eqnarray}
	\nonumber
	\mathbf{B_{\text{dip}}} \left(\mathbf{R_\mu}\right) &=& m_{\rm ord} \cdot \mu_{\text{0}} / 4 \pi \times \\
	\nonumber
	\cos \left(\mathbf{k}\cdot\mathbf{R_\mu}\right) &\cdot& \left[\underbrace{\sum_{\text{i}} \cos\left(\mathbf{k}\cdot\left[\mathbf{r_i}-\mathbf{R_{\text{ref}}}\right]\right)\cdot\left(\frac{3\mathbf{r_i}\cdot(\mathbf{a}\cdot\mathbf{r_i})}{\left|\mathbf{r_i}\right|^{5}}-\frac{\mathbf{a}}{\left|\mathbf{r_i}\right|^{3}}\right) \pm  \sum_{\text{i}} \sin\left(\mathbf{k}\cdot\left[\mathbf{r_i}-\mathbf{R_{\text{ref}}}\right]\right)\cdot\left(\frac{3\mathbf{r_i}\cdot(\mathbf{b}\cdot\mathbf{r_i})}{\left|\mathbf{r_i}\right|^{5}}-\frac{\mathbf{b}}{\left|\mathbf{r_i}\right|^{3}}\right)}_{\mathbf{C_{\text{dip}}}}\right] \\
	\nonumber
	+ \sin \left(\mathbf{k}\cdot\mathbf{R_\mu}\right) &\cdot& \left[\underbrace{\pm \sum_{\text{i}} \sin\left(\mathbf{k}\cdot\left[\mathbf{r_i}-\mathbf{R_{\text{ref}}}\right]\right)\cdot\left(\frac{3\mathbf{r_i}\cdot(\mathbf{a}\cdot\mathbf{r_i})}{\left|\mathbf{r_i}\right|^{5}}-\frac{\mathbf{a}}{\left|\mathbf{r_i}\right|^{3}}\right) \pm \sum_{\text{i}} \cos\left(\mathbf{k}\cdot\left[\mathbf{r_i}-\mathbf{R_{\text{ref}}}\right]\right)\cdot\left(\frac{3\mathbf{r_i}\cdot(\mathbf{b}\cdot\mathbf{r_i})}{\left|\mathbf{r_i}\right|^{5}}-\frac{\mathbf{b}}{\left|\mathbf{r_i}\right|^{3}}\right)}_{\mathbf{S_{\text{dip}}}}\right]\\
	&&
	\label{A4}
\end{eqnarray}
\end{widetext}

Owing to the fact that we are dealing with an incommensurate structure, one can replace the argument $\mathbf{k}\cdot\mathbf{R_\mu}$ by the continuous variable $\delta$, taking all values between $0$ and $2\pi$. The dipolar field at the muon site eventually reads:

\begin{equation}
	\mathbf{B_{\text{dip}}} \left(\mathbf{R_\mu}\right) = \mathbf{C_{\text{dip}}} \cdot \cos \delta + \mathbf{S_{\text{dip}}} \cdot \sin \delta
	\label{A5}	
\end{equation}

The advantage in using this form is that the lattice sums $\mathbf{C_{\text{dip}}}$ and $\mathbf{S_{\text{dip}}}$ in Eq. \ref{A4} need to be computed only once. A direct numerical application will converge to better than 0.1\% within a sphere containing more than $\simeq 10^{5}$ unit cells, or even faster by making use of Ewald's summation method.\\

For computing the \emph{contact field}, we make use of Eq. \ref{M5} from main text. Using Eq. \ref{A2} for describing the magnetic moments of the three Mn ions nearest neighbors of the muon (see also Tab. \ref{tab:musites}), we end up with a similar expression as Eq. \ref{A5}, namely

\begin{equation}
	\mathbf{B_{\text{cont}}} \left(\mathbf{R_\mu}\right) = \left( \mathbf{C_{\text{cont}}} \cdot \cos \delta + \mathbf{S_{\text{cont}}} \cdot \sin \delta \right) \quad ,
	\label{A6}	
\end{equation}

with  

\begin{widetext}
\begin{eqnarray}
	\nonumber
	\mathbf{C_{\text{cont}}} &=& \frac{A_{\rm cont} \cdot m_{\rm ord}}{N} \left[ \sum_{\text{i=1}}^{\text{N}} \cos\left(\mathbf{k}\cdot\left[\mathbf{r_i}-\mathbf{R_{\text{ref}}}\right]\right)\cdot \mathbf{a} \pm \sum_{\text{i=1}}^{\text{N}} \sin\left(\mathbf{k}\cdot\left[\mathbf{r_i}-\mathbf{R_{\text{ref}}}\right]\right)\cdot \mathbf{b} \right]\\
	\mathbf{S_{\text{cont}}} &=& \frac{A_{\rm cont} \cdot m_{\rm ord}}{N} \left[ \pm \sum_{\text{i=1}}^{\text{N}} \sin\left(\mathbf{k}\cdot\left[\mathbf{r_i}-\mathbf{R_{\text{ref}}}\right]\right)\cdot\mathbf{b} \pm \sum_{\text{i=1}}^{\text{N}} \cos\left(\mathbf{k}\cdot\left[\mathbf{r_i}-\mathbf{R_{\text{ref}}}\right]\right)\cdot\mathbf{a} \right] \quad ,
	\label{A7}	
\end{eqnarray}
\end{widetext}

where $A_{\text{cont}}$ is the contact coupling constant. The sums in Eq. \ref{A7} run over the $N$ Mn ions, nearest neighbors of the muon. In the case of MnGe, where both muons and Mn ions are sitting on the 4a site of the space group P$2_{1}3$, $N = 3$. 

Note that $A_{\rm cont}$ is given in T.$\mu_{\rm B}^{-1}$ throughout the paper. For comparison with data reported in Ref. \onlinecite{Amato2014}, it is however possible to express it in mol.emu$^{-1}$ via $A_{\rm cont} [\text{mol.emu}^{-1}] = 10^{4}/\left(\mathcal{N}_{\rm A} \cdot \mu_{\rm B}\right) \cdot A_{\rm cont} [\text{T.}\mu_{\rm B}^{-1}] \simeq 1.791 \cdot A_{\rm cont} [\text{T.}\mu_{\rm B}^{-1}]$, where we have used the Avogadro number $\mathcal{N}_{\rm A} = 6.022 \cdot 10^{23}$ mol$^{-1}$ and the Bohr magneton $\mu_{\rm B} = 9.274 \cdot 10^{-21}$ emu. The factor $10^{4}$ is due to conversion from SI to CGS unit system ({\it i.e.} 1 T = 10$^{4}$ G).

The (total) \emph{local} field will finally be obtained as

\begin{equation}
\begin{array}{r@{}l}
	\mathbf{B_{\text{loc}}} \left(\mathbf{R_\mu}\right) &{}= \mathbf{B_{\text{dip}}} \left(\mathbf{R_\mu}\right)  + \mathbf{B_{\text{cont}}} \left(\mathbf{R_\mu}\right)\\
	&{}= \left(\mathbf{C_{\text{dip}}}+\mathbf{C_{\text{cont}}}\right) \cdot \cos \delta + \left(\mathbf{S_{\text{dip}}}+\mathbf{S_{\text{cont}}}\right) \cdot \sin \delta
	\label{A8}	
\end{array}
\end{equation}

Note that for our numerical calculations, we have sampled $10^{4}$ values of $\delta$, which results in the distribution displayed in Fig. \ref{distrib}.

\section{ab initio calculations}
\label{sec:appendixB}

We have estimated the electronic structure of MnGe with DFT as implemented in the \emph{ab initio} package \textsc{QuantumESPRESSO}\cite{Giannozzi_2009} which uses a plane wave basis set and the pseudopotential approach to remove chemically inactive core electrons from the description.
The Generalized Gradient Approximation\cite{PhysRevLett.77.3865} (GGA) was used to estimate the exchange and correlation potential and the ultrasoft pseudopotentials\cite{PhysRevB.41.7892} of the GBRV library\cite{Garrity_2014} provided an optimal compromise between efficiency and accuracy. The basis set
was expanded up to a kinetic energy cut-off of 70 Ry and 60 Ry for the unit cell and the supercell simulations respectively and up to 500 Ry for the charge density.
The reciprocal space was sampled with a $8 \times 8 \times 8$ Monkhorst-Pack (MP) grid.\cite{PhysRevB.13.5188} The reciprocal space of the supercells containing 32 formula units (f.u.) was sampled with the Baldereshi point\cite{PhysRevB.7.5212} when performing structural relaxations.
A $4 \times 4 \times 4$ MP grid was used when analyzing the effect of the impurity on the magnetic properties of the system (\emph{vide infra}).
The relaxed unit cell obtained with these parameters has a lattice constant of 4.763~{\AA} in very good agreement with experimental estimations. DFT simulations also correctly accounted for the high-spin to low-spin transition discussed in Ref.~\onlinecite{Roessler2012}.
The magnetic moment per f.u. in the high spin configuration is $2.02\,\mu_{\mathrm{B}}$ while it reduces to about $1\,\mu_{\mathrm{B}}$ in the low-spin state, in agreement with previously published results\cite{Roessler2012}.

The identification of the muon site was conducted with the method discussed in Refs.~\onlinecite{jp5125876,1402-4896-88-6-068510,PhysRevB.87.121108,PhysRevB.88.064423}. The muon was modeled as a hydrogen atom. A supercell containing 32 f.u. was used to identify the candidate muon sites which are provided by the structural relaxation of the system containing the impurity.
In order to sample all the interstitial space of the unit cell, as already mentioned in Section \ref{sec:results}, we first setup a $6 \times 6 \times 6$ grid of interstitial positions for the impurity to be used as a starting configuration for the structural relaxation. We later removed the positions that were too close (less than 1~{\AA}) to one of the atoms of the hosting material and we finally got rid of all symmetry equivalent initial positions. This eventually led to a set of 14 initial positions for the impurity.
The location of the atoms and of the impurity were optimized (keeping the lattice parameter fixed) until forces were lower than $0.5 \cdot 10^{-3}$ Ry/Bohr and the total energy difference between self consistent field steps was less than $1\cdot 10^{-4}$ Ry. All 14 structural relaxations converged to one symmetry equivalent position which is described in the text and shown in Fig.~\ref{ab_initio}.

Recent DFT results\cite{PhysRevLett.114.017602} brought back the attention on the ``passivity'' of the muon probe showing that, in some peculiar compounds, the muon can conceal the material response as a result of the perturbation introduced by its positive charge. 
We have verified that this is not the case for MnGe. Indeed, differently from the striking effect of pressure discussed above, in the supercell simulations the muon's perturbation produces detectable structural displacements that, however, are more pronounced for Ge and are always smaller than 0.12~{\AA}\footnote{The displacements obtained from supercell simulations are roughly of the order of magnitude of the accuracy of the calculation}. These small dislocations lead to negligible modifications of the magnetic moment of the Mn atoms surrounding the muon, confirming the validity of $\mu$SR results.

Using the double adiabatic approximation\cite{jp5125876}, we estimated the muon's ground state motion energy and the corresponding wave-function.
We used both Dirichlet and periodic boundary conditions to solve the Schr\"odinger equation for the muon in the potential obtained from the interpolation of a point could surrounding the muon and we obtained, with both approaches, $E_{\mathrm{GS}} = 0.57$~eV.
This result is in agreement with the experimentally observed stable muon site since the energy barrier between neighboring sites is about twice as high.


\begin{thebibliography}{55}
\expandafter\ifx\csname natexlab\endcsname\relax\def\natexlab#1{#1}\fi
\expandafter\ifx\csname bibnamefont\endcsname\relax
  \def\bibnamefont#1{#1}\fi
\expandafter\ifx\csname bibfnamefont\endcsname\relax
  \def\bibfnamefont#1{#1}\fi
\expandafter\ifx\csname citenamefont\endcsname\relax
  \def\citenamefont#1{#1}\fi
\expandafter\ifx\csname url\endcsname\relax
  \def\url#1{\texttt{#1}}\fi
\expandafter\ifx\csname urlprefix\endcsname\relax\def\urlprefix{URL }\fi
\providecommand{\bibinfo}[2]{#2}
\providecommand{\eprint}[2][]{\url{#2}}

\bibitem[{\citenamefont{Moriya and Kawabata}(1973)}]{Moriya1973}
\bibinfo{author}{\bibfnamefont{T.}~\bibnamefont{Moriya}} \bibnamefont{and}
  \bibinfo{author}{\bibfnamefont{A.}~\bibnamefont{Kawabata}},
  \bibinfo{journal}{Journal of the Physical Society of Japan}
  \textbf{\bibinfo{volume}{34}}, \bibinfo{pages}{639} (\bibinfo{year}{1973}).

\bibitem[{\citenamefont{Brazovskii}(1975)}]{Brazovskii1975}
\bibinfo{author}{\bibfnamefont{S.}~\bibnamefont{Brazovskii}},
  \bibinfo{journal}{Zh. Eksp. Teor. Fiz.} \textbf{\bibinfo{volume}{68}},
  \bibinfo{pages}{175} (\bibinfo{year}{1975}).

\bibitem[{\citenamefont{Pfleiderer et~al.}(2004)\citenamefont{Pfleiderer,
  Reznik, Pintschovius, Lohneysen, Garst, and Rosch}}]{Pfleiderer2004}
\bibinfo{author}{\bibfnamefont{C.}~\bibnamefont{Pfleiderer}},
  \bibinfo{author}{\bibfnamefont{D.}~\bibnamefont{Reznik}},
  \bibinfo{author}{\bibfnamefont{L.}~\bibnamefont{Pintschovius}},
  \bibinfo{author}{\bibfnamefont{H.~v.} \bibnamefont{Lohneysen}},
  \bibinfo{author}{\bibfnamefont{M.}~\bibnamefont{Garst}}, \bibnamefont{and}
  \bibinfo{author}{\bibfnamefont{A.}~\bibnamefont{Rosch}},
  \bibinfo{journal}{Nature} \textbf{\bibinfo{volume}{427}},
  \bibinfo{pages}{227} (\bibinfo{year}{2004}).

\bibitem[{\citenamefont{Pedrazzini et~al.}(2007)\citenamefont{Pedrazzini,
  Wilhelm, Jaccard, Jarlborg, Schmidt, Hanfland, Akselrud, Yuan, Schwarz, Grin
  et~al.}}]{Pedrazzini2007}
\bibinfo{author}{\bibfnamefont{P.}~\bibnamefont{Pedrazzini}},
  \bibinfo{author}{\bibfnamefont{H.}~\bibnamefont{Wilhelm}},
  \bibinfo{author}{\bibfnamefont{D.}~\bibnamefont{Jaccard}},
  \bibinfo{author}{\bibfnamefont{T.}~\bibnamefont{Jarlborg}},
  \bibinfo{author}{\bibfnamefont{M.}~\bibnamefont{Schmidt}},
  \bibinfo{author}{\bibfnamefont{M.}~\bibnamefont{Hanfland}},
  \bibinfo{author}{\bibfnamefont{L.}~\bibnamefont{Akselrud}},
  \bibinfo{author}{\bibfnamefont{H.~Q.} \bibnamefont{Yuan}},
  \bibinfo{author}{\bibfnamefont{U.}~\bibnamefont{Schwarz}},
  \bibinfo{author}{\bibfnamefont{Y.}~\bibnamefont{Grin}}, \bibnamefont{et~al.},
  \bibinfo{journal}{Phys. Rev. Lett.} \textbf{\bibinfo{volume}{98}},
  \bibinfo{pages}{047204} (\bibinfo{year}{2007}).

\bibitem[{\citenamefont{Pappas et~al.}(2009)\citenamefont{Pappas,
  Leli\`evre-Berna, Falus, Bentley, Moskvin, Grigoriev, Fouquet, and
  Farago}}]{Pappas2009}
\bibinfo{author}{\bibfnamefont{C.}~\bibnamefont{Pappas}},
  \bibinfo{author}{\bibfnamefont{E.}~\bibnamefont{Leli\`evre-Berna}},
  \bibinfo{author}{\bibfnamefont{P.}~\bibnamefont{Falus}},
  \bibinfo{author}{\bibfnamefont{P.~M.} \bibnamefont{Bentley}},
  \bibinfo{author}{\bibfnamefont{E.}~\bibnamefont{Moskvin}},
  \bibinfo{author}{\bibfnamefont{S.}~\bibnamefont{Grigoriev}},
  \bibinfo{author}{\bibfnamefont{P.}~\bibnamefont{Fouquet}}, \bibnamefont{and}
  \bibinfo{author}{\bibfnamefont{B.}~\bibnamefont{Farago}},
  \bibinfo{journal}{Phys. Rev. Lett.} \textbf{\bibinfo{volume}{102}},
  \bibinfo{pages}{197202} (\bibinfo{year}{2009}).

\bibitem[{\citenamefont{Hamann et~al.}(2011)\citenamefont{Hamann, Lamago, Wolf,
  v.~L\"ohneysen, and Reznik}}]{Hamann2011}
\bibinfo{author}{\bibfnamefont{A.}~\bibnamefont{Hamann}},
  \bibinfo{author}{\bibfnamefont{D.}~\bibnamefont{Lamago}},
  \bibinfo{author}{\bibfnamefont{T.}~\bibnamefont{Wolf}},
  \bibinfo{author}{\bibfnamefont{H.}~\bibnamefont{v.~L\"ohneysen}},
  \bibnamefont{and} \bibinfo{author}{\bibfnamefont{D.}~\bibnamefont{Reznik}},
  \bibinfo{journal}{Phys. Rev. Lett.} \textbf{\bibinfo{volume}{107}},
  \bibinfo{pages}{037207} (\bibinfo{year}{2011}).

\bibitem[{\citenamefont{R\"o\ss{}ler et~al.}(2006)\citenamefont{R\"o\ss{}ler,
  Bogdanov, and Pfleiderer}}]{Roessler2006}
\bibinfo{author}{\bibfnamefont{U.~K.} \bibnamefont{R\"o\ss{}ler}},
  \bibinfo{author}{\bibfnamefont{A.~N.} \bibnamefont{Bogdanov}},
  \bibnamefont{and}
  \bibinfo{author}{\bibfnamefont{C.}~\bibnamefont{Pfleiderer}},
  \bibinfo{journal}{Nature} \textbf{\bibinfo{volume}{442}},
  \bibinfo{pages}{797} (\bibinfo{year}{2006}).

\bibitem[{\citenamefont{Wilhelm et~al.}(2012)\citenamefont{Wilhelm, Baenitz,
  Schmidt, Naylor, Lortz, R\"o\ss{}ler, Leonov, and Bogdanov}}]{Wilhelm2012}
\bibinfo{author}{\bibfnamefont{H.}~\bibnamefont{Wilhelm}},
  \bibinfo{author}{\bibfnamefont{M.}~\bibnamefont{Baenitz}},
  \bibinfo{author}{\bibfnamefont{M.}~\bibnamefont{Schmidt}},
  \bibinfo{author}{\bibfnamefont{C.}~\bibnamefont{Naylor}},
  \bibinfo{author}{\bibfnamefont{R.}~\bibnamefont{Lortz}},
  \bibinfo{author}{\bibfnamefont{U.~K.} \bibnamefont{R\"o\ss{}ler}},
  \bibinfo{author}{\bibfnamefont{A.~A.} \bibnamefont{Leonov}},
  \bibnamefont{and} \bibinfo{author}{\bibfnamefont{A.~N.}
  \bibnamefont{Bogdanov}}, \bibinfo{journal}{Journal of Physics: Condensed
  Matter} \textbf{\bibinfo{volume}{24}}, \bibinfo{pages}{294204}
  (\bibinfo{year}{2012}).

\bibitem[{\citenamefont{M\"uhlbauer et~al.}(2009)\citenamefont{M\"uhlbauer,
  Binz, Jonietz, Pfleiderer, Rosch, Neubauer, Georgii, and
  B\"oni}}]{Muhlbauer2009}
\bibinfo{author}{\bibfnamefont{S.}~\bibnamefont{M\"uhlbauer}},
  \bibinfo{author}{\bibfnamefont{B.}~\bibnamefont{Binz}},
  \bibinfo{author}{\bibfnamefont{F.}~\bibnamefont{Jonietz}},
  \bibinfo{author}{\bibfnamefont{C.}~\bibnamefont{Pfleiderer}},
  \bibinfo{author}{\bibfnamefont{A.}~\bibnamefont{Rosch}},
  \bibinfo{author}{\bibfnamefont{A.}~\bibnamefont{Neubauer}},
  \bibinfo{author}{\bibfnamefont{R.}~\bibnamefont{Georgii}}, \bibnamefont{and}
  \bibinfo{author}{\bibfnamefont{P.}~\bibnamefont{B\"oni}},
  \bibinfo{journal}{Science} \textbf{\bibinfo{volume}{323}},
  \bibinfo{pages}{915} (\bibinfo{year}{2009}).

\bibitem[{\citenamefont{Grigoriev et~al.}(2010)\citenamefont{Grigoriev,
  Maleyev, Moskvin, Dyadkin, Fouquet, and Eckerlebe}}]{Grigoriev2010}
\bibinfo{author}{\bibfnamefont{S.~V.} \bibnamefont{Grigoriev}},
  \bibinfo{author}{\bibfnamefont{S.~V.} \bibnamefont{Maleyev}},
  \bibinfo{author}{\bibfnamefont{E.~V.} \bibnamefont{Moskvin}},
  \bibinfo{author}{\bibfnamefont{V.~A.} \bibnamefont{Dyadkin}},
  \bibinfo{author}{\bibfnamefont{P.}~\bibnamefont{Fouquet}}, \bibnamefont{and}
  \bibinfo{author}{\bibfnamefont{H.}~\bibnamefont{Eckerlebe}},
  \bibinfo{journal}{Phys. Rev. B} \textbf{\bibinfo{volume}{81}},
  \bibinfo{pages}{144413} (\bibinfo{year}{2010}).

\bibitem[{\citenamefont{Janoschek et~al.}(2013)\citenamefont{Janoschek, Garst,
  Bauer, Krautscheid, Georgii, B\"oni, and Pfleiderer}}]{Janoschek2013}
\bibinfo{author}{\bibfnamefont{M.}~\bibnamefont{Janoschek}},
  \bibinfo{author}{\bibfnamefont{M.}~\bibnamefont{Garst}},
  \bibinfo{author}{\bibfnamefont{A.}~\bibnamefont{Bauer}},
  \bibinfo{author}{\bibfnamefont{P.}~\bibnamefont{Krautscheid}},
  \bibinfo{author}{\bibfnamefont{R.}~\bibnamefont{Georgii}},
  \bibinfo{author}{\bibfnamefont{P.}~\bibnamefont{B\"oni}}, \bibnamefont{and}
  \bibinfo{author}{\bibfnamefont{C.}~\bibnamefont{Pfleiderer}},
  \bibinfo{journal}{Phys. Rev. B} \textbf{\bibinfo{volume}{87}},
  \bibinfo{pages}{134407} (\bibinfo{year}{2013}).

\bibitem[{\citenamefont{Yu et~al.}(2010)\citenamefont{Yu, Onose, Kanazawa,
  Park, Han, Matsui, Nagaosa, and Tokura}}]{Yu2010}
\bibinfo{author}{\bibfnamefont{X.~Z.} \bibnamefont{Yu}},
  \bibinfo{author}{\bibfnamefont{Y.}~\bibnamefont{Onose}},
  \bibinfo{author}{\bibfnamefont{N.}~\bibnamefont{Kanazawa}},
  \bibinfo{author}{\bibfnamefont{J.~H.} \bibnamefont{Park}},
  \bibinfo{author}{\bibfnamefont{J.~H.} \bibnamefont{Han}},
  \bibinfo{author}{\bibfnamefont{Y.}~\bibnamefont{Matsui}},
  \bibinfo{author}{\bibfnamefont{N.}~\bibnamefont{Nagaosa}}, \bibnamefont{and}
  \bibinfo{author}{\bibfnamefont{Y.}~\bibnamefont{Tokura}},
  \bibinfo{journal}{Nature} \textbf{\bibinfo{volume}{465}},
  \bibinfo{pages}{901} (\bibinfo{year}{2010}), ISSN \bibinfo{issn}{0028-0836}.

\bibitem[{\citenamefont{Tsvyashchenko}(1984)}]{Tsvyashchenko1984}
\bibinfo{author}{\bibfnamefont{A.}~\bibnamefont{Tsvyashchenko}},
  \bibinfo{journal}{Journal of the Less Common Metals}
  \textbf{\bibinfo{volume}{99}}, \bibinfo{pages}{L9 } (\bibinfo{year}{1984}).

\bibitem[{\citenamefont{Deutsch
  et~al.}(2014{\natexlab{a}})\citenamefont{Deutsch, Makarova, Hansen,
  Fernandez-Diaz, Sidorov, Tsvyashchenko, Fomicheva, Porcher, Petit, Koepernik
  et~al.}}]{Deutsch2014}
\bibinfo{author}{\bibfnamefont{M.}~\bibnamefont{Deutsch}},
  \bibinfo{author}{\bibfnamefont{O.~L.} \bibnamefont{Makarova}},
  \bibinfo{author}{\bibfnamefont{T.~C.} \bibnamefont{Hansen}},
  \bibinfo{author}{\bibfnamefont{M.~T.} \bibnamefont{Fernandez-Diaz}},
  \bibinfo{author}{\bibfnamefont{V.~A.} \bibnamefont{Sidorov}},
  \bibinfo{author}{\bibfnamefont{A.~V.} \bibnamefont{Tsvyashchenko}},
  \bibinfo{author}{\bibfnamefont{L.~N.} \bibnamefont{Fomicheva}},
  \bibinfo{author}{\bibfnamefont{F.}~\bibnamefont{Porcher}},
  \bibinfo{author}{\bibfnamefont{S.}~\bibnamefont{Petit}},
  \bibinfo{author}{\bibfnamefont{K.}~\bibnamefont{Koepernik}},
  \bibnamefont{et~al.}, \bibinfo{journal}{Phys. Rev. B}
  \textbf{\bibinfo{volume}{89}}, \bibinfo{pages}{180407}
  (\bibinfo{year}{2014}{\natexlab{a}}).

\bibitem[{\citenamefont{Makarova et~al.}(2012)\citenamefont{Makarova,
  Tsvyashchenko, Andre, Porcher, Fomicheva, Rey, and Mirebeau}}]{Makarova2012}
\bibinfo{author}{\bibfnamefont{O.~L.} \bibnamefont{Makarova}},
  \bibinfo{author}{\bibfnamefont{A.~V.} \bibnamefont{Tsvyashchenko}},
  \bibinfo{author}{\bibfnamefont{G.}~\bibnamefont{Andre}},
  \bibinfo{author}{\bibfnamefont{F.}~\bibnamefont{Porcher}},
  \bibinfo{author}{\bibfnamefont{L.~N.} \bibnamefont{Fomicheva}},
  \bibinfo{author}{\bibfnamefont{N.}~\bibnamefont{Rey}}, \bibnamefont{and}
  \bibinfo{author}{\bibfnamefont{I.}~\bibnamefont{Mirebeau}},
  \bibinfo{journal}{Phys. Rev. B} \textbf{\bibinfo{volume}{85}},
  \bibinfo{pages}{205205} (\bibinfo{year}{2012}).

\bibitem[{\citenamefont{Kanazawa et~al.}(2011)\citenamefont{Kanazawa, Onose,
  Arima, Okuyama, Ohoyama, Wakimoto, Kakurai, Ishiwata, and
  Tokura}}]{Kanazawa2011}
\bibinfo{author}{\bibfnamefont{N.}~\bibnamefont{Kanazawa}},
  \bibinfo{author}{\bibfnamefont{Y.}~\bibnamefont{Onose}},
  \bibinfo{author}{\bibfnamefont{T.}~\bibnamefont{Arima}},
  \bibinfo{author}{\bibfnamefont{D.}~\bibnamefont{Okuyama}},
  \bibinfo{author}{\bibfnamefont{K.}~\bibnamefont{Ohoyama}},
  \bibinfo{author}{\bibfnamefont{S.}~\bibnamefont{Wakimoto}},
  \bibinfo{author}{\bibfnamefont{K.}~\bibnamefont{Kakurai}},
  \bibinfo{author}{\bibfnamefont{S.}~\bibnamefont{Ishiwata}}, \bibnamefont{and}
  \bibinfo{author}{\bibfnamefont{Y.}~\bibnamefont{Tokura}},
  \bibinfo{journal}{Phys. Rev. Lett.} \textbf{\bibinfo{volume}{106}},
  \bibinfo{pages}{156603} (\bibinfo{year}{2011}).

\bibitem[{\citenamefont{Shiomi et~al.}(2013)\citenamefont{Shiomi, Kanazawa,
  Shibata, Onose, and Tokura}}]{Shiomi2013}
\bibinfo{author}{\bibfnamefont{Y.}~\bibnamefont{Shiomi}},
  \bibinfo{author}{\bibfnamefont{N.}~\bibnamefont{Kanazawa}},
  \bibinfo{author}{\bibfnamefont{K.}~\bibnamefont{Shibata}},
  \bibinfo{author}{\bibfnamefont{Y.}~\bibnamefont{Onose}}, \bibnamefont{and}
  \bibinfo{author}{\bibfnamefont{Y.}~\bibnamefont{Tokura}},
  \bibinfo{journal}{Phys. Rev. B} \textbf{\bibinfo{volume}{88}},
  \bibinfo{pages}{064409} (\bibinfo{year}{2013}).

\bibitem[{\citenamefont{Kanazawa et~al.}(2012)\citenamefont{Kanazawa, Kim,
  Inosov, White, Egetenmeyer, Gavilano, Ishiwata, Onose, Arima, Keimer
  et~al.}}]{Kanazawa2012}
\bibinfo{author}{\bibfnamefont{N.}~\bibnamefont{Kanazawa}},
  \bibinfo{author}{\bibfnamefont{J.-H.} \bibnamefont{Kim}},
  \bibinfo{author}{\bibfnamefont{D.~S.} \bibnamefont{Inosov}},
  \bibinfo{author}{\bibfnamefont{J.~S.} \bibnamefont{White}},
  \bibinfo{author}{\bibfnamefont{N.}~\bibnamefont{Egetenmeyer}},
  \bibinfo{author}{\bibfnamefont{J.~L.} \bibnamefont{Gavilano}},
  \bibinfo{author}{\bibfnamefont{S.}~\bibnamefont{Ishiwata}},
  \bibinfo{author}{\bibfnamefont{Y.}~\bibnamefont{Onose}},
  \bibinfo{author}{\bibfnamefont{T.}~\bibnamefont{Arima}},
  \bibinfo{author}{\bibfnamefont{B.}~\bibnamefont{Keimer}},
  \bibnamefont{et~al.}, \bibinfo{journal}{Phys. Rev. B}
  \textbf{\bibinfo{volume}{86}}, \bibinfo{pages}{134425}
  (\bibinfo{year}{2012}).

\bibitem[{\citenamefont{Deutsch
  et~al.}(2014{\natexlab{b}})\citenamefont{Deutsch, Bonville, Tsvyashchenko,
  Fomicheva, Porcher, Damay, Petit, and Mirebeau}}]{Deutsch2014b}
\bibinfo{author}{\bibfnamefont{M.}~\bibnamefont{Deutsch}},
  \bibinfo{author}{\bibfnamefont{P.}~\bibnamefont{Bonville}},
  \bibinfo{author}{\bibfnamefont{A.~V.} \bibnamefont{Tsvyashchenko}},
  \bibinfo{author}{\bibfnamefont{L.~N.} \bibnamefont{Fomicheva}},
  \bibinfo{author}{\bibfnamefont{F.}~\bibnamefont{Porcher}},
  \bibinfo{author}{\bibfnamefont{F.}~\bibnamefont{Damay}},
  \bibinfo{author}{\bibfnamefont{S.}~\bibnamefont{Petit}}, \bibnamefont{and}
  \bibinfo{author}{\bibfnamefont{I.}~\bibnamefont{Mirebeau}},
  \bibinfo{journal}{Phys. Rev. B} \textbf{\bibinfo{volume}{90}},
  \bibinfo{pages}{144401} (\bibinfo{year}{2014}{\natexlab{b}}).

\bibitem[{\citenamefont{Altynbaev et~al.}(2014)\citenamefont{Altynbaev,
  Siegfried, Dyadkin, Moskvin, Menzel, Heinemann, Dewhurst, Fomicheva,
  Tsvyashchenko, and Grigoriev}}]{Altynbaev2014}
\bibinfo{author}{\bibfnamefont{E.}~\bibnamefont{Altynbaev}},
  \bibinfo{author}{\bibfnamefont{S.-A.} \bibnamefont{Siegfried}},
  \bibinfo{author}{\bibfnamefont{V.}~\bibnamefont{Dyadkin}},
  \bibinfo{author}{\bibfnamefont{E.}~\bibnamefont{Moskvin}},
  \bibinfo{author}{\bibfnamefont{D.}~\bibnamefont{Menzel}},
  \bibinfo{author}{\bibfnamefont{A.}~\bibnamefont{Heinemann}},
  \bibinfo{author}{\bibfnamefont{C.}~\bibnamefont{Dewhurst}},
  \bibinfo{author}{\bibfnamefont{L.}~\bibnamefont{Fomicheva}},
  \bibinfo{author}{\bibfnamefont{A.}~\bibnamefont{Tsvyashchenko}},
  \bibnamefont{and}
  \bibinfo{author}{\bibfnamefont{S.}~\bibnamefont{Grigoriev}},
  \bibinfo{journal}{Phys. Rev. B} \textbf{\bibinfo{volume}{90}},
  \bibinfo{pages}{174420} (\bibinfo{year}{2014}).

\bibitem[{\citenamefont{R\"ossler}(2012)}]{Roessler2012}
\bibinfo{author}{\bibfnamefont{U.~K.} \bibnamefont{R\"ossler}},
  \bibinfo{journal}{Journal of Physics: Conference Series}
  \textbf{\bibinfo{volume}{391}}, \bibinfo{pages}{012104}
  (\bibinfo{year}{2012}).

\bibitem[{\citenamefont{Martin et~al.}(2016)\citenamefont{Martin, Mirebeau,
  Deutsch, Iti\'e, Rueff, R\"ossler, Koepernik, Fomicheva, and
  Tsvyashchenko}}]{Martin2016}
\bibinfo{author}{\bibfnamefont{N.}~\bibnamefont{Martin}},
  \bibinfo{author}{\bibfnamefont{I.}~\bibnamefont{Mirebeau}},
  \bibinfo{author}{\bibfnamefont{M.}~\bibnamefont{Deutsch}},
  \bibinfo{author}{\bibfnamefont{J.-P.} \bibnamefont{Iti\'e}},
  \bibinfo{author}{\bibfnamefont{J.-P.} \bibnamefont{Rueff}},
  \bibinfo{author}{\bibfnamefont{U.}~\bibnamefont{R\"ossler}},
  \bibinfo{author}{\bibfnamefont{K.}~\bibnamefont{Koepernik}},
  \bibinfo{author}{\bibfnamefont{L.}~\bibnamefont{Fomicheva}},
  \bibnamefont{and}
  \bibinfo{author}{\bibfnamefont{A.}~\bibnamefont{Tsvyashchenko}}
  (\bibinfo{year}{2016}), \eprint{arXiv:1601.05332v1}.

\bibitem[{\citenamefont{Amato et~al.}(2014)\citenamefont{Amato, Dalmas~de
  R\'eotier, Andreica, Yaouanc, Suter, Lapertot, Pop, Morenzoni, Bonf\`a,
  Bernardini et~al.}}]{Amato2014}
\bibinfo{author}{\bibfnamefont{A.}~\bibnamefont{Amato}},
  \bibinfo{author}{\bibfnamefont{P.}~\bibnamefont{Dalmas~de R\'eotier}},
  \bibinfo{author}{\bibfnamefont{D.}~\bibnamefont{Andreica}},
  \bibinfo{author}{\bibfnamefont{A.}~\bibnamefont{Yaouanc}},
  \bibinfo{author}{\bibfnamefont{A.}~\bibnamefont{Suter}},
  \bibinfo{author}{\bibfnamefont{G.}~\bibnamefont{Lapertot}},
  \bibinfo{author}{\bibfnamefont{I.~M.} \bibnamefont{Pop}},
  \bibinfo{author}{\bibfnamefont{E.}~\bibnamefont{Morenzoni}},
  \bibinfo{author}{\bibfnamefont{P.}~\bibnamefont{Bonf\`a}},
  \bibinfo{author}{\bibfnamefont{F.}~\bibnamefont{Bernardini}},
  \bibnamefont{et~al.}, \bibinfo{journal}{Phys. Rev. B}
  \textbf{\bibinfo{volume}{89}}, \bibinfo{pages}{184425}
  (\bibinfo{year}{2014}).

\bibitem[{\citenamefont{Hayano et~al.}(1979)\citenamefont{Hayano, Uemura,
  Imazato, Nishida, Yamazaki, and Kubo}}]{Hayano1979}
\bibinfo{author}{\bibfnamefont{R.~S.} \bibnamefont{Hayano}},
  \bibinfo{author}{\bibfnamefont{Y.~J.} \bibnamefont{Uemura}},
  \bibinfo{author}{\bibfnamefont{J.}~\bibnamefont{Imazato}},
  \bibinfo{author}{\bibfnamefont{N.}~\bibnamefont{Nishida}},
  \bibinfo{author}{\bibfnamefont{T.}~\bibnamefont{Yamazaki}}, \bibnamefont{and}
  \bibinfo{author}{\bibfnamefont{R.}~\bibnamefont{Kubo}},
  \bibinfo{journal}{Phys. Rev. B} \textbf{\bibinfo{volume}{20}},
  \bibinfo{pages}{850} (\bibinfo{year}{1979}).

\bibitem[{\citenamefont{Grigoriev et~al.}(2013)\citenamefont{Grigoriev,
  Potapova, Siegfried, Dyadkin, Moskvin, Dmitriev, Menzel, Dewhurst,
  Chernyshov, Sadykov et~al.}}]{Grigoriev2013}
\bibinfo{author}{\bibfnamefont{S.~V.} \bibnamefont{Grigoriev}},
  \bibinfo{author}{\bibfnamefont{N.~M.} \bibnamefont{Potapova}},
  \bibinfo{author}{\bibfnamefont{S.-A.} \bibnamefont{Siegfried}},
  \bibinfo{author}{\bibfnamefont{V.~A.} \bibnamefont{Dyadkin}},
  \bibinfo{author}{\bibfnamefont{E.~V.} \bibnamefont{Moskvin}},
  \bibinfo{author}{\bibfnamefont{V.}~\bibnamefont{Dmitriev}},
  \bibinfo{author}{\bibfnamefont{D.}~\bibnamefont{Menzel}},
  \bibinfo{author}{\bibfnamefont{C.~D.} \bibnamefont{Dewhurst}},
  \bibinfo{author}{\bibfnamefont{D.}~\bibnamefont{Chernyshov}},
  \bibinfo{author}{\bibfnamefont{R.~A.} \bibnamefont{Sadykov}},
  \bibnamefont{et~al.}, \bibinfo{journal}{Phys. Rev. Lett.}
  \textbf{\bibinfo{volume}{110}}, \bibinfo{pages}{207201}
  (\bibinfo{year}{2013}).

\bibitem[{\citenamefont{Chizhikov and Dmitrienko}(2012)}]{Chizhikov2012}
\bibinfo{author}{\bibfnamefont{V.~A.} \bibnamefont{Chizhikov}}
  \bibnamefont{and} \bibinfo{author}{\bibfnamefont{V.~E.}
  \bibnamefont{Dmitrienko}}, \bibinfo{journal}{Phys. Rev. B}
  \textbf{\bibinfo{volume}{85}}, \bibinfo{pages}{014421}
  (\bibinfo{year}{2012}).

\bibitem[{\citenamefont{Dmitrienko and Chizhikov}(2012)}]{Dmitrienko2012}
\bibinfo{author}{\bibfnamefont{V.~E.} \bibnamefont{Dmitrienko}}
  \bibnamefont{and} \bibinfo{author}{\bibfnamefont{V.~A.}
  \bibnamefont{Chizhikov}}, \bibinfo{journal}{Phys. Rev. Lett.}
  \textbf{\bibinfo{volume}{108}}, \bibinfo{pages}{187203}
  (\bibinfo{year}{2012}).

\bibitem[{\citenamefont{Bonf\`a et~al.}(2015)\citenamefont{Bonf\`a, Sartori,
  and Renzi}}]{Bonfa2015}
\bibinfo{author}{\bibfnamefont{P.}~\bibnamefont{Bonf\`a}},
  \bibinfo{author}{\bibfnamefont{F.}~\bibnamefont{Sartori}}, \bibnamefont{and}
  \bibinfo{author}{\bibfnamefont{R.~D.} \bibnamefont{Renzi}},
  \bibinfo{journal}{The Journal of Physical Chemistry C}
  \textbf{\bibinfo{volume}{119}}, \bibinfo{pages}{4278} (\bibinfo{year}{2015}).

\bibitem[{\citenamefont{Campbell et~al.}(1994)\citenamefont{Campbell, Amato,
  Gygax, Herlach, Schenck, Cywinski, and Kilcoyne}}]{Campbell1994}
\bibinfo{author}{\bibfnamefont{I.~A.} \bibnamefont{Campbell}},
  \bibinfo{author}{\bibfnamefont{A.}~\bibnamefont{Amato}},
  \bibinfo{author}{\bibfnamefont{F.~N.} \bibnamefont{Gygax}},
  \bibinfo{author}{\bibfnamefont{D.}~\bibnamefont{Herlach}},
  \bibinfo{author}{\bibfnamefont{A.}~\bibnamefont{Schenck}},
  \bibinfo{author}{\bibfnamefont{R.}~\bibnamefont{Cywinski}}, \bibnamefont{and}
  \bibinfo{author}{\bibfnamefont{S.~H.} \bibnamefont{Kilcoyne}},
  \bibinfo{journal}{Phys. Rev. Lett.} \textbf{\bibinfo{volume}{72}},
  \bibinfo{pages}{1291} (\bibinfo{year}{1994}).

\bibitem[{\citenamefont{Keren et~al.}(1996)\citenamefont{Keren, Mendels,
  Campbell, and Lord}}]{Keren1996}
\bibinfo{author}{\bibfnamefont{A.}~\bibnamefont{Keren}},
  \bibinfo{author}{\bibfnamefont{P.}~\bibnamefont{Mendels}},
  \bibinfo{author}{\bibfnamefont{I.~A.} \bibnamefont{Campbell}},
  \bibnamefont{and} \bibinfo{author}{\bibfnamefont{J.}~\bibnamefont{Lord}},
  \bibinfo{journal}{Phys. Rev. Lett.} \textbf{\bibinfo{volume}{77}},
  \bibinfo{pages}{1386} (\bibinfo{year}{1996}).

\bibitem[{\citenamefont{Pickup et~al.}(2009)\citenamefont{Pickup, Cywinski,
  Pappas, Farago, and Fouquet}}]{Pickup2009}
\bibinfo{author}{\bibfnamefont{R.~M.} \bibnamefont{Pickup}},
  \bibinfo{author}{\bibfnamefont{R.}~\bibnamefont{Cywinski}},
  \bibinfo{author}{\bibfnamefont{C.}~\bibnamefont{Pappas}},
  \bibinfo{author}{\bibfnamefont{B.}~\bibnamefont{Farago}}, \bibnamefont{and}
  \bibinfo{author}{\bibfnamefont{P.}~\bibnamefont{Fouquet}},
  \bibinfo{journal}{Phys. Rev. Lett.} \textbf{\bibinfo{volume}{102}},
  \bibinfo{pages}{097202} (\bibinfo{year}{2009}).

\bibitem[{\citenamefont{Bewley and Cywinski}(1998)}]{Bewley1998}
\bibinfo{author}{\bibfnamefont{R.~I.} \bibnamefont{Bewley}} \bibnamefont{and}
  \bibinfo{author}{\bibfnamefont{R.}~\bibnamefont{Cywinski}},
  \bibinfo{journal}{Phys. Rev. B} \textbf{\bibinfo{volume}{58}},
  \bibinfo{pages}{11544} (\bibinfo{year}{1998}).

\bibitem[{\citenamefont{Jackson et~al.}(2000)\citenamefont{Jackson, Binns,
  Forgan, Morenzoni, Niedermayer, Gl\"uckler, Hofer, Luetkens, Prokscha,
  Riseman et~al.}}]{Jackson2000}
\bibinfo{author}{\bibfnamefont{T.~J.} \bibnamefont{Jackson}},
  \bibinfo{author}{\bibfnamefont{C.}~\bibnamefont{Binns}},
  \bibinfo{author}{\bibfnamefont{E.~M.} \bibnamefont{Forgan}},
  \bibinfo{author}{\bibfnamefont{E.}~\bibnamefont{Morenzoni}},
  \bibinfo{author}{\bibfnamefont{C.}~\bibnamefont{Niedermayer}},
  \bibinfo{author}{\bibfnamefont{H.}~\bibnamefont{Gl\"uckler}},
  \bibinfo{author}{\bibfnamefont{A.}~\bibnamefont{Hofer}},
  \bibinfo{author}{\bibfnamefont{H.}~\bibnamefont{Luetkens}},
  \bibinfo{author}{\bibfnamefont{T.}~\bibnamefont{Prokscha}},
  \bibinfo{author}{\bibfnamefont{T.~M.} \bibnamefont{Riseman}},
  \bibnamefont{et~al.}, \bibinfo{journal}{Journal of Physics: Condensed Matter}
  \textbf{\bibinfo{volume}{12}}, \bibinfo{pages}{1399} (\bibinfo{year}{2000}).

\bibitem[{\citenamefont{Bert et~al.}(2006)\citenamefont{Bert, Mendels, Olariu,
  Blanchard, Collin, Amato, Baines, and Hillier}}]{Bert2006}
\bibinfo{author}{\bibfnamefont{F.}~\bibnamefont{Bert}},
  \bibinfo{author}{\bibfnamefont{P.}~\bibnamefont{Mendels}},
  \bibinfo{author}{\bibfnamefont{A.}~\bibnamefont{Olariu}},
  \bibinfo{author}{\bibfnamefont{N.}~\bibnamefont{Blanchard}},
  \bibinfo{author}{\bibfnamefont{G.}~\bibnamefont{Collin}},
  \bibinfo{author}{\bibfnamefont{A.}~\bibnamefont{Amato}},
  \bibinfo{author}{\bibfnamefont{C.}~\bibnamefont{Baines}}, \bibnamefont{and}
  \bibinfo{author}{\bibfnamefont{A.~D.} \bibnamefont{Hillier}},
  \bibinfo{journal}{Phys. Rev. Lett.} \textbf{\bibinfo{volume}{97}},
  \bibinfo{pages}{117203} (\bibinfo{year}{2006}).

\bibitem[{\citenamefont{Dalmas~de R\'eotier
  et~al.}(2006)\citenamefont{Dalmas~de R\'eotier, Yaouanc, Keller, Cervellino,
  Roessli, Baines, Forget, Vaju, Gubbens, Amato et~al.}}]{Dalmas2006}
\bibinfo{author}{\bibfnamefont{P.}~\bibnamefont{Dalmas~de R\'eotier}},
  \bibinfo{author}{\bibfnamefont{A.}~\bibnamefont{Yaouanc}},
  \bibinfo{author}{\bibfnamefont{L.}~\bibnamefont{Keller}},
  \bibinfo{author}{\bibfnamefont{A.}~\bibnamefont{Cervellino}},
  \bibinfo{author}{\bibfnamefont{B.}~\bibnamefont{Roessli}},
  \bibinfo{author}{\bibfnamefont{C.}~\bibnamefont{Baines}},
  \bibinfo{author}{\bibfnamefont{A.}~\bibnamefont{Forget}},
  \bibinfo{author}{\bibfnamefont{C.}~\bibnamefont{Vaju}},
  \bibinfo{author}{\bibfnamefont{P.~C.~M.} \bibnamefont{Gubbens}},
  \bibinfo{author}{\bibfnamefont{A.}~\bibnamefont{Amato}},
  \bibnamefont{et~al.}, \bibinfo{journal}{Phys. Rev. Lett.}
  \textbf{\bibinfo{volume}{96}}, \bibinfo{pages}{127202}
  (\bibinfo{year}{2006}).

\bibitem[{\citenamefont{Guo et~al.}(2014)\citenamefont{Guo, Xing, Tong, Tao,
  Watanabe, and an~Xu}}]{Guo2014}
\bibinfo{author}{\bibfnamefont{H.}~\bibnamefont{Guo}},
  \bibinfo{author}{\bibfnamefont{H.}~\bibnamefont{Xing}},
  \bibinfo{author}{\bibfnamefont{J.}~\bibnamefont{Tong}},
  \bibinfo{author}{\bibfnamefont{Q.}~\bibnamefont{Tao}},
  \bibinfo{author}{\bibfnamefont{I.}~\bibnamefont{Watanabe}}, \bibnamefont{and}
  \bibinfo{author}{\bibfnamefont{Z.}~\bibnamefont{an~Xu}},
  \bibinfo{journal}{Journal of Physics: Condensed Matter}
  \textbf{\bibinfo{volume}{26}}, \bibinfo{pages}{436002}
  (\bibinfo{year}{2014}).

\bibitem[{\citenamefont{de~R\'eotier and Yaouanc}(1997)}]{Dalmas1997}
\bibinfo{author}{\bibfnamefont{P.~D.} \bibnamefont{de~R\'eotier}}
  \bibnamefont{and} \bibinfo{author}{\bibfnamefont{A.}~\bibnamefont{Yaouanc}},
  \bibinfo{journal}{Journal of Physics: Condensed Matter}
  \textbf{\bibinfo{volume}{9}}, \bibinfo{pages}{9113} (\bibinfo{year}{1997}).

\bibitem[{\citenamefont{Johnston}(2006)}]{Johnston2006}
\bibinfo{author}{\bibfnamefont{D.~C.} \bibnamefont{Johnston}},
  \bibinfo{journal}{Phys. Rev. B} \textbf{\bibinfo{volume}{74}},
  \bibinfo{pages}{184430} (\bibinfo{year}{2006}).

\bibitem[{\citenamefont{Maisuradze et~al.}(2011)\citenamefont{Maisuradze,
  Guguchia, Graneli, R\o{}nnow, Berger, and Keller}}]{Maisuradze2011}
\bibinfo{author}{\bibfnamefont{A.}~\bibnamefont{Maisuradze}},
  \bibinfo{author}{\bibfnamefont{Z.}~\bibnamefont{Guguchia}},
  \bibinfo{author}{\bibfnamefont{B.}~\bibnamefont{Graneli}},
  \bibinfo{author}{\bibfnamefont{H.~M.} \bibnamefont{R\o{}nnow}},
  \bibinfo{author}{\bibfnamefont{H.}~\bibnamefont{Berger}}, \bibnamefont{and}
  \bibinfo{author}{\bibfnamefont{H.}~\bibnamefont{Keller}},
  \bibinfo{journal}{Phys. Rev. B} \textbf{\bibinfo{volume}{84}},
  \bibinfo{pages}{064433} (\bibinfo{year}{2011}).

\bibitem[{\citenamefont{Lancaster et~al.}(2015)\citenamefont{Lancaster,
  Williams, Thomas, Xiao, Pratt, Blundell, Loudon, Hesjedal, Clark, Hatton
  et~al.}}]{Lancaster2015}
\bibinfo{author}{\bibfnamefont{T.}~\bibnamefont{Lancaster}},
  \bibinfo{author}{\bibfnamefont{R.~C.} \bibnamefont{Williams}},
  \bibinfo{author}{\bibfnamefont{I.~O.} \bibnamefont{Thomas}},
  \bibinfo{author}{\bibfnamefont{F.}~\bibnamefont{Xiao}},
  \bibinfo{author}{\bibfnamefont{F.~L.} \bibnamefont{Pratt}},
  \bibinfo{author}{\bibfnamefont{S.~J.} \bibnamefont{Blundell}},
  \bibinfo{author}{\bibfnamefont{J.~C.} \bibnamefont{Loudon}},
  \bibinfo{author}{\bibfnamefont{T.}~\bibnamefont{Hesjedal}},
  \bibinfo{author}{\bibfnamefont{S.~J.} \bibnamefont{Clark}},
  \bibinfo{author}{\bibfnamefont{P.~D.} \bibnamefont{Hatton}},
  \bibnamefont{et~al.}, \bibinfo{journal}{Phys. Rev. B}
  \textbf{\bibinfo{volume}{91}}, \bibinfo{pages}{224408}
  (\bibinfo{year}{2015}).

\bibitem[{\citenamefont{Storchak et~al.}(2011)\citenamefont{Storchak, Brewer,
  Lichti, Lograsso, and Schlagel}}]{Storchak2011}
\bibinfo{author}{\bibfnamefont{V.~G.} \bibnamefont{Storchak}},
  \bibinfo{author}{\bibfnamefont{J.~H.} \bibnamefont{Brewer}},
  \bibinfo{author}{\bibfnamefont{R.~L.} \bibnamefont{Lichti}},
  \bibinfo{author}{\bibfnamefont{T.~A.} \bibnamefont{Lograsso}},
  \bibnamefont{and} \bibinfo{author}{\bibfnamefont{D.~L.}
  \bibnamefont{Schlagel}}, \bibinfo{journal}{Phys. Rev. B}
  \textbf{\bibinfo{volume}{83}}, \bibinfo{pages}{140404}
  (\bibinfo{year}{2011}).

\bibitem[{\citenamefont{Andreica}(2001)}]{Andreica2001}
\bibinfo{author}{\bibfnamefont{D.-A.} \bibnamefont{Andreica}}, Ph.D. thesis,
  \bibinfo{school}{ETH Zurich} (\bibinfo{year}{2001}).

\bibitem[{\citenamefont{Schenck et~al.}(2001)\citenamefont{Schenck, Andreica,
  Gygax, and Ott}}]{Schenck2001}
\bibinfo{author}{\bibfnamefont{A.}~\bibnamefont{Schenck}},
  \bibinfo{author}{\bibfnamefont{D.}~\bibnamefont{Andreica}},
  \bibinfo{author}{\bibfnamefont{F.~N.} \bibnamefont{Gygax}}, \bibnamefont{and}
  \bibinfo{author}{\bibfnamefont{H.~R.} \bibnamefont{Ott}},
  \bibinfo{journal}{Phys. Rev. B} \textbf{\bibinfo{volume}{65}},
  \bibinfo{pages}{024444} (\bibinfo{year}{2001}).

\bibitem[{\citenamefont{Schenck et~al.}(2003)\citenamefont{Schenck, Gygax, and
  \ifmmode~\bar{O}\else \={O}\fi{}nuki}}]{Schenck2003}
\bibinfo{author}{\bibfnamefont{A.}~\bibnamefont{Schenck}},
  \bibinfo{author}{\bibfnamefont{F.~N.} \bibnamefont{Gygax}}, \bibnamefont{and}
  \bibinfo{author}{\bibfnamefont{Y.}~\bibnamefont{\ifmmode~\bar{O}\else
  \={O}\fi{}nuki}}, \bibinfo{journal}{Phys. Rev. B}
  \textbf{\bibinfo{volume}{68}}, \bibinfo{pages}{104422}
  (\bibinfo{year}{2003}).

\bibitem[{\citenamefont{Giannozzi et~al.}(2009)\citenamefont{Giannozzi, Baroni,
  Bonini, Calandra, Car, Cavazzoni, Ceresoli, Chiarotti, Cococcioni, Dabo
  et~al.}}]{Giannozzi_2009}
\bibinfo{author}{\bibfnamefont{P.}~\bibnamefont{Giannozzi}},
  \bibinfo{author}{\bibfnamefont{S.}~\bibnamefont{Baroni}},
  \bibinfo{author}{\bibfnamefont{N.}~\bibnamefont{Bonini}},
  \bibinfo{author}{\bibfnamefont{M.}~\bibnamefont{Calandra}},
  \bibinfo{author}{\bibfnamefont{R.}~\bibnamefont{Car}},
  \bibinfo{author}{\bibfnamefont{C.}~\bibnamefont{Cavazzoni}},
  \bibinfo{author}{\bibfnamefont{D.}~\bibnamefont{Ceresoli}},
  \bibinfo{author}{\bibfnamefont{G.~L.} \bibnamefont{Chiarotti}},
  \bibinfo{author}{\bibfnamefont{M.}~\bibnamefont{Cococcioni}},
  \bibinfo{author}{\bibfnamefont{I.}~\bibnamefont{Dabo}}, \bibnamefont{et~al.},
  \bibinfo{journal}{J. Phys.: Condens. Matter} \textbf{\bibinfo{volume}{21}},
  \bibinfo{pages}{395502} (\bibinfo{year}{2009}).

\bibitem[{\citenamefont{Perdew et~al.}(1996)\citenamefont{Perdew, Burke, and
  Ernzerhof}}]{PhysRevLett.77.3865}
\bibinfo{author}{\bibfnamefont{J.~P.} \bibnamefont{Perdew}},
  \bibinfo{author}{\bibfnamefont{K.}~\bibnamefont{Burke}}, \bibnamefont{and}
  \bibinfo{author}{\bibfnamefont{M.}~\bibnamefont{Ernzerhof}},
  \bibinfo{journal}{Phys. Rev. Lett.} \textbf{\bibinfo{volume}{77}},
  \bibinfo{pages}{3865} (\bibinfo{year}{1996}).

\bibitem[{\citenamefont{Vanderbilt}(1990)}]{PhysRevB.41.7892}
\bibinfo{author}{\bibfnamefont{D.}~\bibnamefont{Vanderbilt}},
  \bibinfo{journal}{Phys. Rev. B} \textbf{\bibinfo{volume}{41}},
  \bibinfo{pages}{7892} (\bibinfo{year}{1990}).

\bibitem[{\citenamefont{Garrity et~al.}(2014)\citenamefont{Garrity, Bennett,
  Rabe, and Vanderbilt}}]{Garrity_2014}
\bibinfo{author}{\bibfnamefont{K.~F.} \bibnamefont{Garrity}},
  \bibinfo{author}{\bibfnamefont{J.~W.} \bibnamefont{Bennett}},
  \bibinfo{author}{\bibfnamefont{K.~M.} \bibnamefont{Rabe}}, \bibnamefont{and}
  \bibinfo{author}{\bibfnamefont{D.}~\bibnamefont{Vanderbilt}},
  \bibinfo{journal}{Computational Materials Science}
  \textbf{\bibinfo{volume}{81}}, \bibinfo{pages}{446} (\bibinfo{year}{2014}).

\bibitem[{\citenamefont{Monkhorst and Pack}(1976)}]{PhysRevB.13.5188}
\bibinfo{author}{\bibfnamefont{H.~J.} \bibnamefont{Monkhorst}}
  \bibnamefont{and} \bibinfo{author}{\bibfnamefont{J.~D.} \bibnamefont{Pack}},
  \bibinfo{journal}{Phys. Rev. B} \textbf{\bibinfo{volume}{13}},
  \bibinfo{pages}{5188} (\bibinfo{year}{1976}).

\bibitem[{\citenamefont{Baldereschi}(1973)}]{PhysRevB.7.5212}
\bibinfo{author}{\bibfnamefont{A.}~\bibnamefont{Baldereschi}},
  \bibinfo{journal}{Phys. Rev. B} \textbf{\bibinfo{volume}{7}},
  \bibinfo{pages}{5212} (\bibinfo{year}{1973}).

\bibitem[{\citenamefont{Bonf\`a et~al.}(2015)\citenamefont{Bonf\`a, Sartori,
  and Renzi}}]{jp5125876}
\bibinfo{author}{\bibfnamefont{P.}~\bibnamefont{Bonf\`a}},
  \bibinfo{author}{\bibfnamefont{F.}~\bibnamefont{Sartori}}, \bibnamefont{and}
  \bibinfo{author}{\bibfnamefont{R.~D.} \bibnamefont{Renzi}},
  \bibinfo{journal}{The Journal of Physical Chemistry C}
  \textbf{\bibinfo{volume}{119}}, \bibinfo{pages}{4278} (\bibinfo{year}{2015}).

\bibitem[{\citenamefont{M\"oller
  et~al.}(2013{\natexlab{a}})\citenamefont{M\"oller, Bonf\`a, Ceresoli,
  Bernardini, Blundell, Lancaster, Renzi, Marzari, Watanabe, Sulaiman
  et~al.}}]{1402-4896-88-6-068510}
\bibinfo{author}{\bibfnamefont{J.~S.} \bibnamefont{M\"oller}},
  \bibinfo{author}{\bibfnamefont{P.}~\bibnamefont{Bonf\`a}},
  \bibinfo{author}{\bibfnamefont{D.}~\bibnamefont{Ceresoli}},
  \bibinfo{author}{\bibfnamefont{F.}~\bibnamefont{Bernardini}},
  \bibinfo{author}{\bibfnamefont{S.~J.} \bibnamefont{Blundell}},
  \bibinfo{author}{\bibfnamefont{T.}~\bibnamefont{Lancaster}},
  \bibinfo{author}{\bibfnamefont{R.~D.} \bibnamefont{Renzi}},
  \bibinfo{author}{\bibfnamefont{N.}~\bibnamefont{Marzari}},
  \bibinfo{author}{\bibfnamefont{I.}~\bibnamefont{Watanabe}},
  \bibinfo{author}{\bibfnamefont{S.}~\bibnamefont{Sulaiman}},
  \bibnamefont{et~al.}, \bibinfo{journal}{Physica Scripta}
  \textbf{\bibinfo{volume}{88}}, \bibinfo{pages}{068510}
  (\bibinfo{year}{2013}{\natexlab{a}}).

\bibitem[{\citenamefont{M\"oller
  et~al.}(2013{\natexlab{b}})\citenamefont{M\"oller, Ceresoli, Lancaster,
  Marzari, and Blundell}}]{PhysRevB.87.121108}
\bibinfo{author}{\bibfnamefont{J.~S.} \bibnamefont{M\"oller}},
  \bibinfo{author}{\bibfnamefont{D.}~\bibnamefont{Ceresoli}},
  \bibinfo{author}{\bibfnamefont{T.}~\bibnamefont{Lancaster}},
  \bibinfo{author}{\bibfnamefont{N.}~\bibnamefont{Marzari}}, \bibnamefont{and}
  \bibinfo{author}{\bibfnamefont{S.~J.} \bibnamefont{Blundell}},
  \bibinfo{journal}{Phys. Rev. B} \textbf{\bibinfo{volume}{87}},
  \bibinfo{pages}{121108} (\bibinfo{year}{2013}{\natexlab{b}}).

\bibitem[{\citenamefont{Blundell et~al.}(2013)\citenamefont{Blundell, M\"oller,
  Lancaster, Baker, Pratt, Seber, and Lahti}}]{PhysRevB.88.064423}
\bibinfo{author}{\bibfnamefont{S.~J.} \bibnamefont{Blundell}},
  \bibinfo{author}{\bibfnamefont{J.~S.} \bibnamefont{M\"oller}},
  \bibinfo{author}{\bibfnamefont{T.}~\bibnamefont{Lancaster}},
  \bibinfo{author}{\bibfnamefont{P.~J.} \bibnamefont{Baker}},
  \bibinfo{author}{\bibfnamefont{F.~L.} \bibnamefont{Pratt}},
  \bibinfo{author}{\bibfnamefont{G.}~\bibnamefont{Seber}}, \bibnamefont{and}
  \bibinfo{author}{\bibfnamefont{P.~M.} \bibnamefont{Lahti}},
  \bibinfo{journal}{Phys. Rev. B} \textbf{\bibinfo{volume}{88}},
  \bibinfo{pages}{064423} (\bibinfo{year}{2013}).

\bibitem[{\citenamefont{Foronda et~al.}(2015)\citenamefont{Foronda, Lang,
  M\"oller, Lancaster, Boothroyd, Pratt, Giblin, Prabhakaran, and
  Blundell}}]{PhysRevLett.114.017602}
\bibinfo{author}{\bibfnamefont{F.~R.} \bibnamefont{Foronda}},
  \bibinfo{author}{\bibfnamefont{F.}~\bibnamefont{Lang}},
  \bibinfo{author}{\bibfnamefont{J.~S.} \bibnamefont{M\"oller}},
  \bibinfo{author}{\bibfnamefont{T.}~\bibnamefont{Lancaster}},
  \bibinfo{author}{\bibfnamefont{A.~T.} \bibnamefont{Boothroyd}},
  \bibinfo{author}{\bibfnamefont{F.~L.} \bibnamefont{Pratt}},
  \bibinfo{author}{\bibfnamefont{S.~R.} \bibnamefont{Giblin}},
  \bibinfo{author}{\bibfnamefont{D.}~\bibnamefont{Prabhakaran}},
  \bibnamefont{and} \bibinfo{author}{\bibfnamefont{S.~J.}
  \bibnamefont{Blundell}}, \bibinfo{journal}{Phys. Rev. Lett.}
  \textbf{\bibinfo{volume}{114}}, \bibinfo{pages}{017602}
  (\bibinfo{year}{2015}).

\end{thebibliography}
\end{document}